\documentclass[preprint,12pt,authoryear]{elsarticle}

\usepackage[T1]{fontenc}
\usepackage[utf8]{inputenc}
\usepackage{amsmath,amssymb,bm}
\usepackage{graphicx}
\usepackage{booktabs}
\usepackage{multirow}
\usepackage{hyperref}
\hypersetup{colorlinks=true, linkcolor=blue, citecolor=blue, urlcolor=blue}
\usepackage{lineno}

\newcommand{\NuBaselineRBCMean}{2.48}

\newcommand{\NuSingleGRUMean}{2.17}
\newcommand{\NuSingleGRUStd}{0.08}
\newcommand{\NuSingleGRUReductPct}{12.5}
\newcommand{\NuMultiGRUMean}{1.85}
\newcommand{\NuMultiGRUStd}{0.07}
\newcommand{\NuMultiGRUReductPct}{25.5}
\newcommand{\NuSingleNoGRUMean}{2.06}
\newcommand{\NuSingleNoGRUStd}{0.17}
\newcommand{\NuSingleNoGRUReductPct}{17.1}
\newcommand{\NuMultiNoGRUMean}{1.83}
\newcommand{\NuMultiNoGRUStd}{0.00}
\newcommand{\NuMultiNoGRUReductPct}{26.3}
\newcommand{\NuBaseMean}{10.44}
\newcommand{\NuBaseStd}{2.54}
\newcommand{\NuCtrlMean}{12.44}
\newcommand{\NuCtrlStd}{2.41}
\newcommand{\NuImprPct}{19.1}
\newcommand{\ShBaseMean}{198.97}
\newcommand{\ShBaseStd}{44.16}
\newcommand{\ShCtrlMean}{191.93}
\newcommand{\ShCtrlStd}{45.47}
\newcommand{\ShImprPct}{-3.5}
\newcommand{\SvarBaseMean}{0.0629}
\newcommand{\SvarCtrlMean}{0.0497}
\newcommand{\SvarReductPct}{21.0}
\newcommand{\SvarHalfLifeBase}{96.0}
\newcommand{\SvarHalfLifeCtrl}{83.2}
\newcommand{\SvarSpeedup}{1.2}

\journal{International Journal of Heat and Fluid Flow}

\begin{document}

\begin{frontmatter}
\title{Deep reinforcement learning with spatial and temporal awareness for active boundary control of buoyancy-driven convection}

\author[a]{Giorgio Maria Cavallazzi\corref{cor}}
\author[a]{Miguel P\'erez Cuadrado}
\author[a]{Alfredo Pinelli}
\cortext[cor]{Corresponding author.}
\affiliation[a]{organization={Department of Engineering, City St George's -- University of London},
               addressline={Northampton Square},
               city={London},
               postcode={EC1V 0HB},
               country={UK}}
\begin{abstract}
Deep reinforcement learning (DRL) applied to thermal convection control
consistently produces \textit{degenerate actuation}: wall-temperature
policies whose outputs are saturated, pseudo-random, or spatially
incoherent.
Two compounding deficiencies are responsible: multilayer-perceptron
policies that discard spatial flow structure, and memoryless policies
that cannot distinguish self-induced flow changes from background
evolution.
Together they degrade the actuation into forms whose relation to the
convective topology cannot be read off and which are not realisable at
the actuator, even when cell coalescence (the merging of convection
rolls into fewer, larger structures), which would reduce $\mathrm{Nu}$,
is accessible to boundary actuation.

The present framework addresses both causes through four targeted
design choices: convolutional policy networks, Gated Recurrent Unit
(GRU) memory, off-policy training (TD3\,/\,MADDPG), and
action-smoothness constraints.
A systematic $2\times 2$ factorial design isolates the contribution of
each component.
On Rayleigh--B\'{e}nard convection at $\mathrm{Ra} = 10{,}000$, all
four configurations achieve cell coalescence and reduce $\mathrm{Nu}$
to as low as $1.83$ ($26\%$ below the uncontrolled baseline) in
350 episodes, without the full-field data augmentation required by
prior work.
Crucially, coalescence is achieved even by the single-agent
configuration, demonstrating that the multi-agent formulation is not a
prerequisite once the policy architecture is sufficiently expressive.
Applied to double-diffusive convection in the salt-finger regime, the
framework spontaneously discovers a travelling-wave actuation whose
phase speed adapts to the evolving mixing state of the flow, enhancing
heat transfer by $\NuImprPct\%$ and reducing salinity variance by
$\SvarReductPct\%$.
\end{abstract}

\begin{keyword}
Active heat and flow control \sep deep reinforcement learning \sep multi-agent control \sep partial observability \sep  smooth actuation \sep Rayleigh--B\'{e}nard convection \sep
double-diffusive convection \sep cell coalescence 
\end{keyword}

\end{frontmatter}


\section{Introduction}
\label{sec:intro}

Active control of buoyancy-driven flows, where thermal or compositional
density differences drive fluid motion, is a problem of broad scientific
and engineering relevance.
The ability to suppress or enhance convective heat and mass transfer
through boundary actuation has direct applications in thermal
management, geophysical modelling, and industrial processes involving
stratified fluids.
The present work targets two such mechanisms --- cell coalescence in
Rayleigh--B\'{e}nard convection and salt-finger disruption in
double-diffusive convection --- and uses deep reinforcement learning
as the means of inducing them at the boundary.
Among buoyancy-driven flows, Rayleigh--B\'{e}nard convection
(RBC), the motion driven by heating a fluid layer from below and
cooling it from above, occupies a central and canonical place.
At moderate Rayleigh numbers ($\mathrm{Ra} = g\,\beta\,\Delta T\,H^3/\nu\,\kappa$,
where $g$ is the gravitational acceleration, $\beta$ the thermal
expansion coefficient, $\Delta T$ the imposed temperature difference,
$H$ the layer height, $\nu$ the kinematic viscosity, and $\kappa$ the
thermal diffusivity) the flow self-organises into steady convection
rolls whose number and spatial arrangement set the global heat transfer
rate, quantified by the Nusselt number ($\mathrm{Nu} = q_w H /
\lambda\,\Delta T$, where $q_w$ is the wall heat flux and $\lambda$
the thermal conductivity).
Reducing $\mathrm{Nu}$ through boundary actuation is physically
equivalent to reorganising this roll topology: by promoting
\textit{cell coalescence} (i.e. the merging of multiple rolls into fewer,
larger structures) the boundary-layer contact area per roll decreases,
plume impingement weakens, and heat transport approaches the conductive
limit.
The roll topology and the coherent structures that organise it ---
plumes, thermals, and a large-scale circulation --- have been
characterised in detail in moderately wide cylindrical
domains~\citep{Sakievich2016}.
The simplicity of the control metric (a single scalar, $\mathrm{Nu}$)
and the geometric interpretability of the underlying mechanism make RBC
a uniquely clean testbed for data-driven control.

However, deriving effective control laws for such flows is fundamentally
difficult: the governing equations are nonlinear, thermal perturbations
at the wall must first diffuse and then be advected through the bulk
before their effect becomes observable, and the optimal strategy is
rarely known \textit{a priori}.
Classical feedback approaches, ranging from proportional--derivative
(PD) controllers to linear optimal control, require detailed knowledge
of the linearised flow dynamics and lose effectiveness as the Rayleigh
number increases and nonlinear interactions intensify~\citep{Jiren2025}.
Data-driven and machine learning methods offer an alternative
route~\citep{Brunton2020,VinuesaBrunton2022}, and reinforcement
learning in particular is well suited to sequential decision-making in
systems where no explicit model is available~\citep{Garnier2021}.

The application of deep reinforcement learning (DRL) to active flow
control was demonstrated by \citet{Rabault2019}, who showed that a
neural network policy trained through interaction with a high-fidelity
numerical solver can discover effective drag-reduction strategies for
flow past a cylinder at $\mathrm{Re} = 100$.
Their work established the paradigm of coupling a DRL agent to a fluid
solver through a standardised environment interface, and has since been
extended across a wide range of configurations.
For turbulent channel flow, where classical active control strategies
rely on wall blowing and suction to suppress near-wall turbulence and
reduce skin-friction drag, \citet{Guastoni2023} demonstrated that
off-policy algorithms can learn effective actuation strategies directly
from DNS data, without requiring a prescribed control law.
In a similar physical setting, \citet{Cavallazzi2025} targeted the
self-sustaining wall regeneration cycle, achieving drag reduction
through stabilisation of velocity streaks.
\citet{Font2025} demonstrated DRL-based control of a 3D turbulent
separation bubble over a flat plate under adverse pressure gradient,
achieving a 9\% reduction in bubble area; the learned policy remained
effective when transferred from the coarse training grid to a finer
resolution.

When the flow domain is wide or the actuation is spatially distributed,
however, a single agent observing the full field becomes increasingly
impractical, both in terms of observation dimensionality and the
difficulty of learning coordinated spatial actuation.
Building on these single-agent results, multi-agent formulations have
been applied to flows past three-dimensional cylinders in the
transitional regime with Reynolds numbers up to
$\mathrm{Re}_D = 400$~\citep{Suarez2025a}.
In the fully turbulent
regime at $\mathrm{Re}_D = 3900$, \citet{Suarez2025b} have discovered
multi-frequency control laws two orders of magnitude more
mass-efficient than conventional periodic forcing.
Most recently, \citet{Garcia2025} have extended this lineage to
airfoil separation control, achieving a $43.9\%$ drag reduction on a
NACA~0012 at $\mathrm{Re}_D = 3000$ via PPO-controlled blowing and
suction at the surface.

In parallel with these aerodynamic applications, a separate thread
has applied DRL to forced-convection cooling. \citet{Wang2023} use a
deep Q-network to regulate inlet conditions in an open cavity with
discrete heat sources, reaching wall-temperature reductions of
approximately $8\,\mathrm{K}$ over manually tuned controllers;
\citet{Wang2024} subsequently show that policies trained on a $2$D
configuration transfer to $3$D at roughly $10\%$ of the direct
training cost, while noting that the translational-invariance
argument behind multi-agent RBC control does not extend cleanly to
forced-convection settings.

A common thread runs through these developments. Given enough
interaction and a large enough network, a reinforcement learning agent
should in principle be able to learn a control law for any of these
problems, so the practical obstacle is not what the policy class can
represent but what the optimiser can find: when a system carries strong
underlying structure and the learning problem is formulated without
reference to it, the region of the action space that has to be explored
grows with the number of actuators and the exploration cost becomes
prohibitive. \citet{Belus2019} made this argument directly for the
stabilisation of a falling liquid film, where the combined
combinatorial size of the output domain grows as a power of the number
of jets, and showed that sharing one policy across local agents and
rewarding each of them locally recovers what a monolithic formulation
loses. That argument, rather than any property of a particular
algorithm, is what the multi-agent formulations of \citet{Vignon2023},
\citet{Font2025} and \citet{Suarez2025b} exploit, and it admits more
than one implementation. \citet{Jeon2025} revisit the configuration of
\citet{Vignon2023} and build the local reflection and rotation
symmetries of the flow into group-invariant convolutional networks,
retaining the multi-agent re-centring for the global translational
invariance and improving the reproducibility of the learnt policies.
The same reasoning applies on the temporal side, where
\citet{WangJFM2024} show for cylinder wake control with sparse pressure
sensing that a policy acting on the instantaneous sensor signal is
handicapped by the delay between actuation and response, and recover
the missing information by lifting the measurement with its time
derivative. Recurrence is an alternative route to that end, and the
contribution of the present work is to supply the spatial and the
temporal structure together, through a convolutional policy with gated
recurrent memory that requires neither re-centring nor a delay horizon
fixed in advance. The two routes to temporal context are compared
directly in Section~\ref{sec:training}.

Returning to RBC, \citet{Beintema2020} were the first to apply RL to
this problem, suppressing convection entirely for
$\mathrm{Ra} \lesssim 3 \times 10^4$ in a unit-aspect-ratio domain and
outperforming classical PD control.
However, their limited domain constrained the flow to a single-cell
configuration.
\citet{Vignon2023} extended the approach to wider periodic domains
using a multi-agent RL (MARL) framework with full-field re-centred
observations, where optimising for $\mathrm{Nu}$ reduction leads to
cell coalescence, reducing $\mathrm{Nu}$ from approximately 2.7 to 2.0
at $\mathrm{Ra} = 10{,}000$.
Their approach draws two distinct benefits from the multi-agent
decomposition: each environment step yields as many transitions as
there are agents, and the re-centring of the full field about each
agent is what allows a single shared policy to act differently on
different segments.
The MARL framework was subsequently extended to three-dimensional RBC
by \citet{Vasanth2025}, achieving convection-intensity reductions of
23.5\% and 8.7\% at $\mathrm{Ra} = 500$ and 750, respectively.
More recent work has extended DRL-based RBC control toward higher
Rayleigh numbers: \citet{Markmann2025} report $\mathrm{Nu}$ reductions
of up to 33\% at moderate $\mathrm{Ra}$ and 10\% in the chaotic regime
using single-agent PPO with reward shaping; \citet{Chen2025} couple RL
with a reduced-order model derived from proper orthogonal decomposition
and autoencoders, achieving $16$--$23\%$ $\mathrm{Nu}$ reduction under
both single- and dual-boundary actuation.

Outside the buoyancy-driven setting, \citet{Holme2026} address the
multi-timescale nature of rotating detonation engine control through
a moving reference frame that separates fast wave propagation from
slower mode transitions, and through a physically anchored
credit-assignment horizon $\gamma = 1 - \Delta t / T_h$ with $T_h$
chosen on the gain-recovery timescale; the action period $\Delta t$
itself is treated as a swept hyperparameter rather than tied to a
particular physical scale.

Despite these advances, the policies reported in the RBC control
literature exhibit a recurring and underappreciated pathology:
\textit{degenerate actuation}.
The actor outputs typically either saturate at the control bounds or
settle into spatially disordered patterns with no clear connection to
the convective topology. Such strategies may lower the reward
functional by a small margin yet encode no physical mechanism.
In turbulent flow control, where the actuation cadence is fast relative
to the flow dynamics, this degeneracy manifests most visibly as
\textit{bang-bang} actuation: policies that alternate rapidly between
extreme actuator values at each step, producing physically
non-realisable switching that is energetically prohibitive and harmful
to actuator hardware.
In RBC at moderate Rayleigh numbers, where the actuation timescale is
comparable to the convective turnover, the same root causes produce a
related but distinct failure mode: spatially incoherent or saturated
wall-temperature patterns that happen to perturb the flow without
driving meaningful roll reorganisation.

We identify two compounding deficiencies as responsible.
The first is \textit{insufficient network expressivity}: all prior DRL
approaches to RBC use MLP policies that receive the flow state as a
flattened vector, discarding spatial locality and translational
structure.
An MLP cannot easily learn that adjacent wall segments should be
actuated in concert to match the wavelength of the underlying
convection rolls, and instead converges to the degenerate patterns
described above.
The second deficiency is the absence of \textit{temporal context}.
In a multi-agent setting, where each agent observes only a local
fraction of the flow domain, the individual observation is no longer a
sufficient statistic for the full system state: the problem is formally
a decentralised partially observable Markov decision process
(Dec-POMDP).
A memoryless policy operating on a single snapshot cannot distinguish
whether a local temperature change is the result of its own prior
actuation or of the natural flow evolution, and this ambiguity drives
the optimiser toward saturated or random outputs as a hedging strategy.
An alternative remedy, recently applied to bluff-body wake control by
\citet{JiaXu2025}, augments the observation with a Kalman-filtered
state estimate rather than introducing recurrence; the recurrent
route adopted here learns the relevant summary internally and scales
more directly to the high-dimensional spatially distributed
observations of multi-agent wall control.
To the best of our knowledge, both deficiencies are present, to a
varying degree, in all existing DRL approaches to thermal convection
control, and they have not been addressed together.

Both deficiencies have known remedies in the broader reinforcement
learning literature that have not yet been brought to bear on
thermal convection control.
A recurrent policy, in particular one built around a gated unit such
as the GRU, carries a hidden state across decision steps that lets
each agent follow the delayed flow response to its own actuation and
separate self-induced changes from background evolution; the
architecture has a long-standing role in partially observable
RL~\citep{Hochreiter1997,Cho2014,Kapturowski2019} and has been shown
to encode genuine flow physics in turbulent wall
modelling~\citep{Bae2022}.
A convolutional encoder, in turn, treats the flow field as a
structured spatial array and extracts coherent features without the
full-field data augmentation of \citet{Vignon2023}; graph neural
networks extend the same inductive bias to arbitrary
geometries~\citep{Kurz2025}, and explainability analyses confirm that
spatially aware policies recover physically meaningful flow
structures~\citep{Beneitez2025}.
Off-policy training closes the loop: TD3 (Twin Delayed Deep
Deterministic policy gradient) for the single-agent case and MADDPG
(Multi-Agent Deep Deterministic Policy Gradient;
\citealt{Lowe2017}) for the multi-agent case reuse past transitions
through a replay buffer, cut the number of costly solver
episodes~\citep{Guastoni2023}, and accommodate recurrent actors
through sequence sampling~\citep{Kapturowski2019}.
%

The present work addresses these gaps simultaneously through four
design choices.
Convolutional policy networks operating on local spatial patches
replace the global MLP observations of \citet{Vignon2023}, providing
spatial feature extraction through local receptive fields and weight
sharing across the domain.
Recurrent actors, augmented with a Gated Recurrent Unit (GRU), maintain
temporal context across actuation steps, enabling each agent to track
the flow response to its own past actions.
Off-policy training, using a deterministic policy gradient algorithm
for the single-agent and a multi-agent extension thereof~\citep{Lowe2017}
for the multi-agent configuration, replaces on-policy methods, storing
past transitions in a replay buffer and reusing them across multiple
updates to reduce the number of expensive solver episodes required.
Finally, action constraints, including a zero-mean projection,
an amplitude cap, and spatial and temporal smoothness penalties,
explicitly discourage the saturated and incoherent outputs that
characterise degenerate policies.

Four configurations are compared: single-agent and multi-agent
formulations, each trained with and without GRU recurrence, allowing
the contributions of temporal memory and of the multi-agent structure
to be assessed independently.
The key finding is that, with sufficient architectural expressivity
and temporal memory, even a single-agent policy achieves cell
coalescence and meaningful $\mathrm{Nu}$ reduction at
$\mathrm{Ra} = 10{,}000$ in a wide periodic domain.
The barrier in prior work was not the agent topology but the policy
architecture, and the multi-agent formulation of \citet{Vignon2023}
is not a prerequisite for discovering the correct physical mechanism.
The multi-agent configuration provides additional benefits, including
better spectral alignment with the dominant convective mode and a
larger $\mathrm{Nu}$ reduction, while GRU recurrence accelerates
convergence by approximately 100 episodes across all configurations.

The framework is validated on two buoyancy-driven configurations of
increasing complexity: RBC at $\mathrm{Ra} = 10{,}000$, where the
objective is to reduce $\mathrm{Nu}$ through cell coalescence, and
double-diffusive convection in the salt-finger regime, where the
objective is to enhance heat transfer and accelerate scalar mixing.
In both cases the learned strategies are smooth, spatially structured,
and physically interpretable, with no trace of the degenerate outputs
that motivated the architectural choices above.

The remainder of the paper is structured as follows.
Section~\ref{sec:physics} describes the governing equations and flow
configurations. Section~\ref{sec:numerics} details the numerical
solver. Section~\ref{sec:drl} presents the DRL formulation.
Sections~\ref{sec:training} and~\ref{sec:evaluation} report training
and evaluation results for the RBC case.
Section~\ref{sec:ddc} presents the double-diffusive extension.
Section~\ref{sec:summary} summarises the main findings.

\section{Physical problem and discrete formulation}
\label{sec:physics}

\subsection{Governing equations, computational domain and boundary conditions}
\label{eq:eqs}

We consider a two-dimensional rectangular domain of width $L_x = 2\pi$
and height $L_y = 2H$, with $H = 1$ the half-height used for
non-dimensionalisation.  The fluid is incompressible and obeys the
Boussinesq approximation: density variations appear only in the buoyancy
term.

Using the free-fall scaling, lengths are made non-dimensional by $H$,
velocities by the free-fall velocity $U_f = \sqrt{g\,\beta\,\Delta T\,H}$,
time by $H / U_f$, pressure by $\rho_0 U_f^2$, and the temperature by
$\Delta T$,  i.e. the difference between the mean
bottom-wall temperature $T_H$ and the top-wall temperature $T_C$.
The resulting dimensionless system reads:
%
%
\begin{align}
  \nabla \cdot \bm{u} &= 0,
  \label{eq:continuity} \\[6pt]
  \frac{\partial \bm{u}}{\partial t}
    + (\bm{u} \cdot \nabla)\bm{u}
  &= -\nabla p
    + \sqrt{\frac{\mathrm{Pr}}{\mathrm{Ra}}}\;\nabla^{2}\,\bm{u}
    + T\,\hat{\bm{e}}_y,
  \label{eq:momentum} \\[6pt]
  \frac{\partial T}{\partial t}
    + (\bm{u} \cdot \nabla)T
  &= \frac{1}{\sqrt{\mathrm{Ra}\,\mathrm{Pr}}}\;\nabla^{2}\, T.
  \label{eq:temperature}
\end{align}

The system is governed by the Rayleigh number $\mathrm{Ra}$ and the
Prandtl number $\mathrm{Pr} = \nu / \kappa_T$. 
No-slip conditions ($\bm{u} = 0$) are imposed at both horizontal walls.
The top wall is held at fixed temperature $T_C = 1$, while the bottom
wall temperature $T_H = 2$ serves as the reference value around which
the agents apply zero-mean perturbations $\delta T(x)$, so that the
effective bottom-wall temperature is $T_H + \delta T(x)$ at any given
time. The $x$-direction is periodic. The non-dimensional temperature is
shifted rather than centred about its mean, so the uniform part of the
buoyancy term in equation~\eqref{eq:momentum} is constant and is
absorbed into the pressure gradient.

Throughout the paper the Nusselt number is evaluated from the wall
temperature gradients as the mean of the values at the two horizontal
walls,
\begin{equation}
  \mathrm{Nu} = \tfrac{1}{2}\bigl(\mathrm{Nu}_{\text{bot}}
    + \mathrm{Nu}_{\text{top}}\bigr),
  \qquad
  \mathrm{Nu}_{\text{wall}} = -\,\frac{H}{T_H - T_C}
    \left\langle \frac{\partial T}{\partial y} \right\rangle_{x,\,\text{wall}},
  \label{eq:nusselt}
\end{equation}
where $\langle \cdot \rangle_{x,\,\text{wall}}$ denotes the average
along the wall. At the bottom wall the gradient is evaluated through the
ghost cell, so that the instantaneous actuated temperature
$T_H + \delta T(x)$ enters the estimate, while the normalisation retains
the fixed reference difference $T_H - T_C$ rather than the instantaneous
wall temperature. Equation~\eqref{eq:nusselt} is the quantity reported
in all figures and tables, and it is the one entering the reward of
Section~\ref{sec:reward}.

All the present simulations use $\mathrm{Ra} = 10{,}000$,
$\mathrm{Pr} = 0.71$ (air) and $L_x = 2\pi$. The uncontrolled steady
state exhibits four counter-rotating convection rolls, yielding a
baseline Nusselt number $\mathrm{Nu}_0 \approx 2.48$. The DRL control
objective is to reduce $\mathrm{Nu}$ by promoting \emph{cell
coalescence}, the merging of rolls into fewer, larger structures under
wall actuation, which weakens convective mixing and drives
$\mathrm{Nu}$ toward the conductive limit ($\mathrm{Nu} = 1$), where
heat transfer occurs entirely by diffusion.

\subsection{Numerical method}
\label{sec:numerics}

The governing equations are solved using a staggered-grid
pressure correction method~\citep{VanKan1986} implemented entirely
in PyTorch, enabling execution on both CPU and GPU.
To avoid numerical dissipation, all spatial operators are discretised
using second-order central differences.
Time integration uses an explicit Adams-Bashforth scheme with adaptive
time stepping controlled by CFL and Fourier number constraints,
maintaining $\mathrm{CFL} \approx 0.1$ and $\mathrm{Fo} < 0.25$
on a $96 \times 64$ uniform grid.
The solver is wrapped in Gymnasium (single-agent) and PettingZoo
(multi-agent) environment interfaces that handle observation
extraction, action application, reward computation, and episode
management.

\section{Deep reinforcement learning formulation}
\label{sec:drl}

In the reinforcement learning paradigm, a decision-making \emph{agent}
interacts repeatedly with an \emph{environment} (here the Boussinesq
flow solver) by observing its current state, selecting an action, and
receiving a scalar reward that reflects how desirable the outcome was.
Through trial and error the agent learns a \emph{policy}, i.e.\ a
mapping from observed states to actions, that maximises the cumulative
reward over time.  Applied to the present problem, the agent observes
the instantaneous temperature and velocity fields, perturbs the
bottom-wall temperature over a set of discrete segments, and is
rewarded according to the resulting change in heat transfer.

Formally, the control task is cast as a Markov decision process
$(\mathcal{S}, \mathcal{A}, P, r, \gamma)$, where $\mathcal{S}$ is the
space of instantaneous flow fields $s_t$,
$a_t \in \mathcal{A} \subseteq
\mathbb{R}^N$ the wall-temperature perturbations over $N$ segments, $P$ the transition kernel
$P(s_{t+1} \mid s_t, a_t)$ (i.e. \ the probability of the flow advancing to
state $s_{t+1}$ from state $s_t$ under actuation $a_t$) governed by the
Boussinesq equations, $r(s_t, a_t)$ the scalar reward, and $\gamma
\in (0,1)$  a discount factor that down-weights rewards received further
in the future, ensuring the cumulative sum remains bounded.
The agent seeks a deterministic policy $\pi_\theta : \mathcal{S} \to \mathcal{A}$
that maximises the cumulative discounted reward over a trajectory,
starting from the cached uncontrolled steady state $s_0$,
\begin{equation}
  J(\theta) \;=\; \sum_{t=0}^{\infty} \gamma^t\, r\bigl(s_t,\,\pi_\theta(s_t)\bigr),
  \qquad s_{t+1} = P\bigl(s_t,\,\pi_\theta(s_t)\bigr),
  \label{eq:return}
\end{equation}
where the flow state $s_{t+1}$ is obtained by advancing the Boussinesq
solver one actuation step from $s_t$ under the wall temperature prescribed
by $\pi_\theta(s_t)$.
In plain terms, the agent is rewarded at every step
for reducing heat transfer, but rewards collected further in the future
count for less (through the factor $\gamma^t$); the policy $\pi_\theta$
is then the wall-temperature strategy that accumulates the highest total
score over the entire episode.

Two off-policy algorithms are used to optimise $J(\theta)$: Twin Delayed Deep Deterministic policy
gradient (TD3; \citealt{Fujimoto2018}) for the single-agent case, and
Multi-Agent Deep Deterministic Policy Gradient (MADDPG; \citealt{Lowe2017})
for the multi-agent case.  
Both TD3 and MADDPG improve the policy by following the gradient of
$J(\theta)$ with respect to the network parameters $\theta$
\citep{Silver2014,Lillicrap2016},
\begin{equation}
  \nabla_\theta J(\theta) \;=\;
  \mathbb{E}\!\left[\,
    \nabla_\theta \pi_\theta(s)\;
    \nabla_a Q^\pi(s, a)\big|_{a = \pi_\theta(s)}
  \right],
  \label{eq:dpg}
\end{equation}
where $Q^\pi(s, a)$ is the action-value function, i.e.\ the total
discounted reward expected from state $s$ when action $a$ is applied
and the policy $\pi_\theta$ is followed thereafter,
\begin{equation}
  Q^\pi(s, a) \;=\; \sum_{t \ge 0} \gamma^t\, r(s_t, a_t),
  \qquad s_0 = s,\; a_0 = a.
  \label{eq:qfunc}
\end{equation}
In practice, $Q^\pi$ is not known and is approximated by a second neural
network (the \emph{critic}), trained concurrently with the policy
(the \emph{actor}). The gradient in equation~\eqref{eq:dpg} then tells
the actor how to adjust its output so that the critic predicts a higher
future reward.

\begin{figure}[htbp]
\centering
\includegraphics[width=0.9\textwidth]{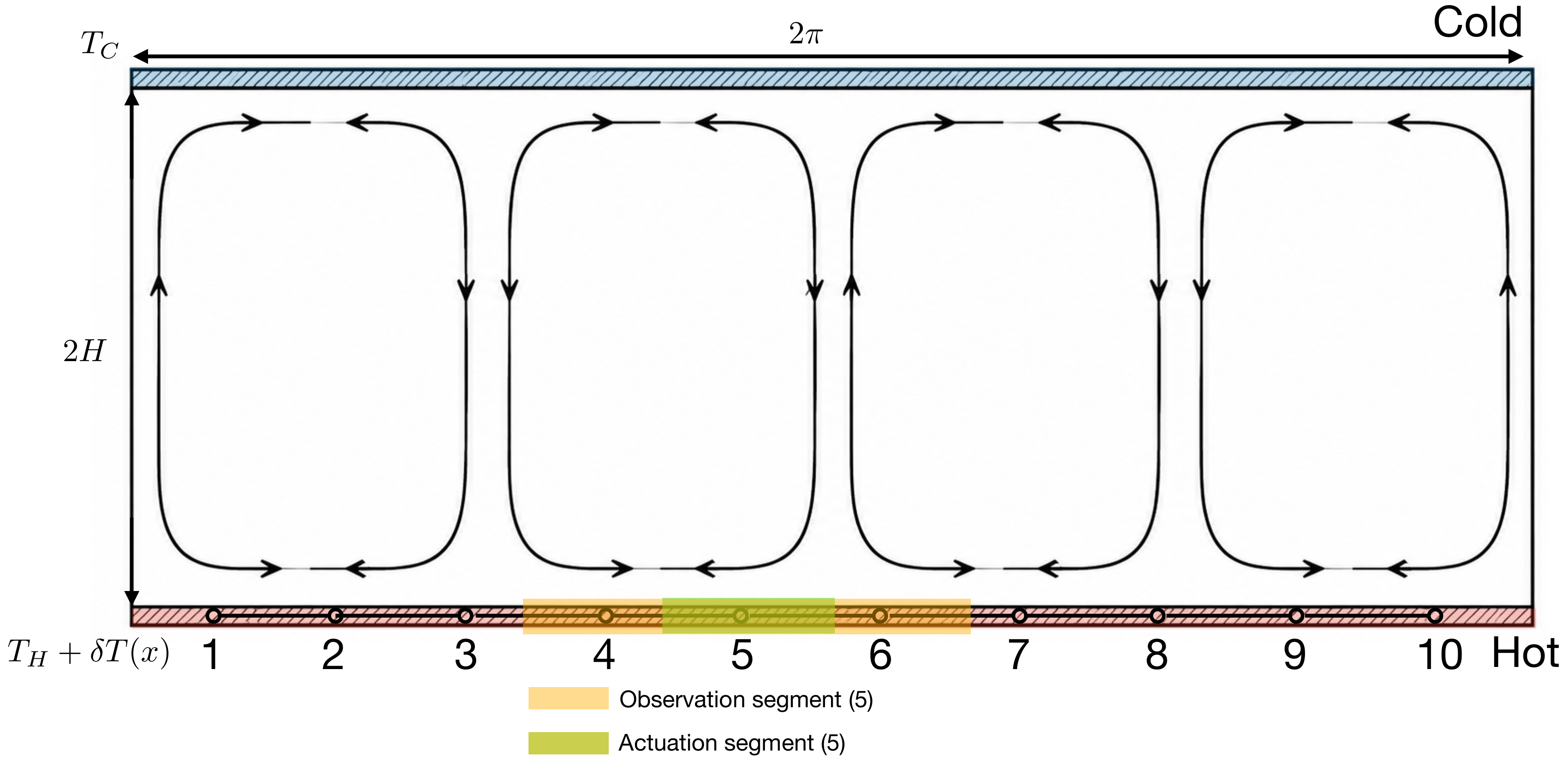}
\caption{Schematic of the control setup for the multi-agent
formulation. The bottom wall is divided into $N = 10$ segments;
agent~5 observes the local flow state over a patch of three
consecutive segments (shaded orange box) but applies its temperature
perturbation $\delta T$ to its own central segment only (shaded green).
The top wall is held at fixed temperature $T_C$; the $x$-direction
is periodic.}
\label{fig:drl_setup}
\end{figure}

When multiple agents control different portions of the wall
simultaneously, each one can only sense its local neighbourhood rather
than the full flow domain.  This limited observability means the
single-agent MDP framework is no longer sufficient: 
the action of one agent affects
the observations of its neighbours, so the agents must learn to
coordinate through the shared reward signal alone, without any direct
communication between them.
%
When multiple agents control different portions of the wall
simultaneously, each one can only sense its local neighbourhood rather
than the full flow domain. The action of one agent therefore affects
the observations of its neighbours, so the agents must learn to
coordinate through the shared reward signal alone, without any direct
communication between them. The multi-agent setting is then described
by extending the MDP tuple to include a separate observation space
$\mathcal{O}_j$ for each agent $j$,
\begin{equation}
  \bigl(\mathcal{S},\;\{\mathcal{A}_j\}_{j=1}^{N},\;P,\;r,\;
    \{\mathcal{O}_j\}_{j=1}^{N},\;\gamma\bigr),
  \label{eq:decpomdp}
\end{equation}
in which agent $j$ receives only the local observation $o_j = \mathcal{O}_j(s)$,
the flow state projected onto the patch of three consecutive wall
segments visible to that agent (Figure~\ref{fig:drl_setup}), and
outputs $a_j \in \mathcal{A}_j = [-1,1]$.

All agents share the same scalar reward $r$, so they are collectively
driven to improve the global heat transfer objective rather than
optimising their individual segments in isolation. During training, a
shared critic is given access to the full flow state and all agents'
actions, allowing it to assess the global consequence of each agent's
local decision. At deployment, however, each agent acts on its local
observation alone, without knowledge of what its neighbours are doing.
This centralised-training, decentralised-execution strategy (CTDE;
\citealt{Lowe2017}) allows the agents to learn coordinated behaviour
during training while remaining independently deployable.

Four configurations are compared: single-agent (TD3) and multi-agent
(MADDPG), each trained with and without GRU recurrence. The effect of
each choice is isolated by comparing all four combinations, as detailed
in the subsections below.
\subsection{Single-agent TD3}
\label{sec:td3}
The single-agent formulation treats the entire bottom wall as one
control surface, observing the full downsampled flow field and
outputting temperature perturbations for all $N = 10$ segments
simultaneously.
The three physical
fields $(T, u, v)$ are downsampled by a factor of~2 (from
$96 \times 64$ to $48 \times 32$), and an optional fourth channel
encodes the previous action as a spatially constant field, giving an
observation $o \in \mathbb{R}^{C \times 32 \times 48}$ with $C = 3$
or~$4$.

The actor processes this observation through four stride-2
convolutional layers with $3 \times 3$ kernels and LeakyReLU
activation, reducing the spatial dimensions from $32 \times 48$ to
$2 \times 3$ and yielding a 768-dimensional feature vector after
flattening. A fully connected layer with LayerNorm maps this to a
256-dimensional hidden representation, followed by a final linear
layer with $\tanh$ activation producing $a \in [-1,1]^{10}$.

The off-policy algorithm underlying the single-agent formulation is
Twin Delayed DDPG (TD3; \citealt{Fujimoto2018}), chosen for its
sample efficiency and training stability relative to on-policy
alternatives such as PPO. The architecture and training
hyperparameters follow \citet{Fujimoto2018}; the reader is referred
there for full algorithmic details.
%
%
%
%

\subsection{Multi-agent MADDPG}
\label{sec:maddpg}

As in the single-agent formulation, the bottom wall is divided into
$N = 10$ equal segments, but here each segment is controlled by a
dedicated agent~$j$ ($j = 1, \ldots, 10$). Agent~$j$ controls the
temperature over its segment of width $L_x / N$ and observes a local
patch spanning 3~consecutive segments (its own plus one neighbour on
each side, with periodic wrapping --- see Figure~\ref{fig:drl_setup}).
All agents share the same policy network.

The value $N = 10$ follows \citet{Vignon2023} and is retained here so that
the two studies remain directly comparable without the segmentation acting
as a confounding difference. Appendix~\ref{app:nseg} reports what happens
when it is varied.

Each agent uses a compact convolutional network: two convolutional
layers extract spatial features from the local observation, followed
by a fully connected layer that produces a single temperature
perturbation for that agent's segment.

During training, a centralised critic has access to the full flow
state and all agents' actions, allowing it to assess the global
consequence of each local decision. At deployment, only the local
actor is used --- each agent acts on its own observation alone,
without knowledge of what its neighbours are doing, consistent with
the strategy described in Section~\ref{sec:drl}.

\subsection{Action constraints}
\label{sec:actions}

All configurations share the same action-processing pipeline. The raw
actor outputs are first shifted to have zero mean,
\begin{equation}
  a'_j = a_j - \frac{1}{N}\sum_{k=0}^{N-1} a_k,
  \label{eq:zero_mean}
\end{equation}
so that the actuation adds no net heat to the bottom wall. The
corrected actions are then rescaled to prevent any single segment
from saturating the actuator. Whenever the largest absolute value
exceeds a threshold (here $C = 0.75$), all actions are scaled down
proportionally,
\begin{equation}
  \tilde{a}_j = \begin{cases}
    a'_j & \text{if } \|a'\|_\infty \le C, \\[3pt]
    a'_j \cdot C / \|a'\|_\infty & \text{otherwise,}
  \end{cases}
  \label{eq:normalise}
\end{equation}
Finally, cubic Hermite interpolation over 10\% of the segment width
smooths the temperature profile at segment boundaries, yielding the
wall condition $T_{\text{wall}}(x) = T_H + \delta T(x)$.

\subsection{Reward design}
\label{sec:reward}

At each actuation step $t$, each agent $j$ receives a scalar measure
of how much the instantaneous heat transfer (i.e.\ the instantaneous
Nusselt number) has been reduced relative to the uncontrolled
baseline (i.e.\ a reward),
\begin{equation}
  r_j^t = m \Bigl[\, n_{\text{shift}}
    - (1 - \beta)\,\mathrm{Nu}_{\text{inst}}^t
    - \beta\,\mathrm{Nu}_{\text{loc},j}^t \,\Bigr].
  \label{eq:reward}
\end{equation}
In equation~\eqref{eq:reward}, $n_{\text{shift}} = 2.67$ is an offset
chosen so that the reward is positive when $\mathrm{Nu}$ falls below
the uncontrolled baseline, and $m = 1$ is a scaling factor. The small
weight $\beta = 0.0015$ blends a local Nusselt number
$\mathrm{Nu}_{\text{loc},j}^t$, computed over agent $j$'s own
segment, into the global value, giving each agent a localised measure
of its individual contribution to the overall heat transfer reduction.

For the single-agent case the local term vanishes ($\beta = 0$) and
the reward simplifies to $r = m(n_{\text{shift}} -
\mathrm{Nu}_{\text{inst}})$: i.e. the agent is rewarded in direct
proportion to the reduction in global heat transfer.

\subsection{GRU architecture}
\label{sec:gru}
The flow response to wall actuation is not instantaneous, as convective
structures take time to form, merge, and reorganise after a change in
boundary temperature. As a result, the instantaneous flow snapshot
alone does not carry enough information to determine the best next
action: the same snapshot could arise from a flow that is responding
to a recent actuation or from one that has been evolving freely, and
the two situations call for different responses. Retaining a memory
of past observations resolves this ambiguity, allowing the agent to
track how the flow is evolving over time and to attribute changes in
heat transfer to its own previous actions.
%
To provide this memory, a Gated Recurrent Unit (GRU;
\citealt{Cho2014}) is incorporated into the policy network, between
the convolutional encoder (the CNN that processes the flow snapshot)
and the output layer that produces the wall-temperature action.
At each actuation step, the GRU updates a hidden state vector that
summarises the history of past observations, which is then carried
forward to the next step. The CNN feature vector is first projected
to dimension $h = 64$ via a linear layer, then passed through the
GRU, whose update equations are:
\begin{align}
  r_t &= \sigma(W_r\, z^t + U_r\, h^{t-1} + b_r),
    \label{eq:gru_reset} \\
  \tilde{h}_t &= \tanh(W_h\, z^t + U_h\,(r_t \odot h^{t-1}) + b_h),
    \label{eq:gru_candidate} \\
  g_t &= \sigma(W_g\, z^t + U_g\, h^{t-1} + b_g),
    \label{eq:gru_update} \\
  h^t &= (1 - g_t) \odot h^{t-1} + g_t \odot \tilde{h}_t,
    \label{eq:gru_output}
\end{align}
where $z^t$ is the projected CNN feature vector, $h^t$ the hidden
state carrying memory across actuation steps, $\sigma$ the sigmoid
function, and $\odot$ elementwise multiplication. The reset gate $r_t$ controls how much of the past memory is retained
at each step, while the update gate $g_t$ balances the contribution
of the new flow observation against that memory.

Training with memory requires temporally contiguous data, since the
past observations must be propagated forward in time to build up a
meaningful memory. Subsequences of length $L = 16$ are therefore
sampled from stored episodes. The first $B = 4$ steps of each
subsequence are used solely to warm up the memory before the policy
update is computed, and are not used in the parameter update
\citep{Kapturowski2019}.

\subsection{Training procedure}
\label{sec:procedure}

All four configurations (single-agent TD3 and multi-agent MADDPG,
each with and without GRU recurrence) are trained for 350 episodes
of 200 actions each (150{,}000 solver time steps per episode).
Before training, the solver is advanced for 400 thermal diffusion
times ($t \kappa_T / H^2 = 400$) to reach the uncontrolled steady
state, which is saved and reloaded at the start of each episode.

Key training parameters (shared across all configurations unless noted)
are listed in Table~\ref{tab:hyperparams}.

\begin{table}[t]
\centering
\caption{Training hyperparameters.}
\label{tab:hyperparams}
\small
\begin{tabular}{lll}
  \toprule
  \textbf{Category} & \textbf{Parameter} & \textbf{Value} \\
  \midrule
  \multirow{3}{*}{Physics}
    & Rayleigh number $\mathrm{Ra}$ & $10{,}000$ \\
    & Prandtl number $\mathrm{Pr}$ & $0.71$ \\
    & Grid & $96 \times 64$ \\
  \midrule
  \multirow{4}{*}{Control}
    & Segments $N$ & 10 \\
    & Substeps per action & 750 \\
    & Amplitude cap $C$ & 0.75 \\
    & Smoothing fraction & 10\% \\
  \midrule
  \multirow{3}{*}{Networks}
    & GRU hidden size & 64 \\
    & Actor / critic LR & $10^{-3}$ / $10^{-3}$ \\
    & Target update $\tau$ & 0.005 \\
  \midrule
  \multirow{5}{*}{Training}
    & Max episodes & 350 \\
    & Episode length & 200 actions \\
    & Batch size & 64 (TD3) / 32 (MADDPG) \\
    & Discount $\gamma$ & 0.99 \\
    & Gradient clip & 0.75 \\
  \midrule
  \multirow{2}{*}{Exploration}
    & Initial noise $\sigma_0$ & 0.1 \\
    & Final noise $\sigma_f$ / decay & 0.01 / 100~ep \\
  \bottomrule
\end{tabular}
\end{table}
Left unconstrained, the actor tends to produce actuation patterns that
are spatially discontinuous, temporally erratic, or saturated at the
maximum allowable value, none of which correspond to physically
meaningful control strategies. To discourage these degenerate
behaviours, four penalty terms are added to the training objective,
each targeting a specific undesirable pattern. The first penalises
any residual deviation from zero mean after the projection
\eqref{eq:zero_mean},
%
\begin{equation}
  \mathcal{L}_{\text{zero}} = \bigl\langle \bar{a}^{\,2} \bigr\rangle, \qquad
  \bar{a} = \frac{1}{N}\sum_{j=0}^{N-1} a_j.
  \label{eq:loss_zero}
\end{equation}
The second penalty addresses spatial discontinuities: abrupt jumps in
wall temperature between adjacent segments can produce localised
flow structures at segment boundaries that are consequences of the
discretisation rather than genuine control features,
\begin{equation}
  \mathcal{L}_{\text{smooth}} = \bigl\langle (a_{j+1} - a_j)^2 \bigr\rangle,
  \label{eq:loss_smooth}
\end{equation}
while the temporal smoothness loss penalises rapid changes between consecutive actuation steps,
\begin{equation}
  \mathcal{L}_{\text{temporal}} = \bigl\langle (a_j^t - a_j^{t-1})^2 \bigr\rangle.
  \label{eq:loss_temporal}
\end{equation}
The temporal smoothness penalty follows the CAPS regularisation of
\citet{Mysore2021}, originally introduced for single-agent continuous
control; the spatial smoothness term enforces coordination across
the segmented action vector and has no direct precedent in that line
of work.
Finally, since the actor output is bounded to $[-1, 1]$ by the $\tanh$
activation, a penalty on large action magnitudes discourages the policy
from saturating at these limits, where the gradient vanishes and
learning stalls,
\begin{equation}
  \mathcal{L}_{\text{energy}} = \bigl\langle a_j^{\,2} \bigr\rangle.
  \label{eq:loss_energy}
\end{equation}
The four penalties are combined with the main training objective
through weights $\lambda_{\text{zero}}$, $\lambda_{\text{smooth}}$,
$\lambda_{\text{temporal}}$, and $\lambda_{\text{energy}}$, whose
values differ between the single-agent and multi-agent configurations.
These are listed in Table~\ref{tab:aux_losses}.

\begin{table}[t]
\centering
\caption{Auxiliary loss weights.}
\label{tab:aux_losses}
\small
\begin{tabular}{lcc}
  \toprule
  \textbf{Loss} & \textbf{MADDPG} & \textbf{TD3} \\
  \midrule
  $\lambda_{\text{zero}}$     & 0.1  & 2.0  \\
  $\lambda_{\text{smooth}}$   & 0.05 & 0.02 \\
  $\lambda_{\text{temporal}}$ & 0.025 & 0.025 \\
  $\lambda_{\text{energy}}$   & 0.1  & 0.1  \\
  \bottomrule
\end{tabular}
\end{table}

\section{Training results: Rayleigh--B\'{e}nard convection}
\label{sec:training}

The four configurations compared throughout this section and the next
are collected in Table~\ref{tab:configs}, together with the labels and
colours used to identify them in every figure. The actor sizes are
worth noting, since they show that the recurrent variants are not
simply larger networks: in the single-agent case the GRU actor is the
smaller of the two, because the projection to the hidden state replaces
the wide dense layer that follows the flattened convolutional features.

\begin{table}[htbp]
\centering
\caption{The four control configurations and the labels used to
identify them in the text and in every figure. Blue denotes the
single-agent configurations and orange the multi-agent ones, with dark
shades for the GRU-equipped variants and light shades for the
memoryless ones, and green the frame-stacked variant. Actor sizes are
the number of trainable parameters of the policy network; the
multi-agent actor is shared by all ten agents.}
\label{tab:configs}
\footnotesize
\setlength{\tabcolsep}{3pt}
\begin{tabular}{llclcr}
  \toprule
  \textbf{Label} & \textbf{Algorithm} & \textbf{Agents} &
  \textbf{Observation} & \textbf{Memory} & \textbf{Actor} \\
  \midrule
  Single-agent, no GRU& TD3    & 1  & full field & none          & 441,066 \\
  Single-agent + GRU   & TD3    & 1  & full field & GRU, $h = 64$ & 315,946 \\
  Multi-agent, no GRU  & MADDPG & 10 & 3 segments & none          & 54,609 \\
  Multi-agent + GRU    & MADDPG & 10 & 3 segments & GRU, $h = 64$ & 79,569 \\
  Multi-agent, stack $k = 4$ & MADDPG & 10 & 3 segments & 4 frames & 56,337 \\
  Multi-agent, wide    & MADDPG & 10 & 3 segments & none          & 81,537 \\
  \bottomrule
\end{tabular}
\end{table}

\begin{figure}[htbp]
\centering
\includegraphics[width=0.95\textwidth]{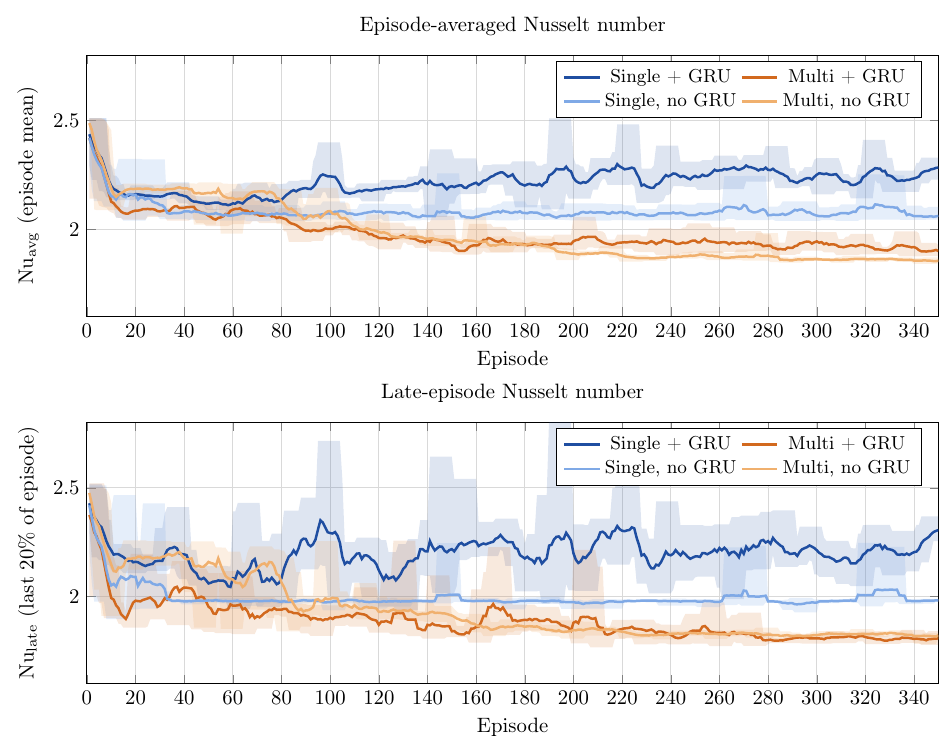}
\caption{Training convergence of the Nusselt number for four agent
configurations.  Top: episode-averaged $\mathrm{Nu}$.
Bottom: late-episode $\mathrm{Nu}$ (last 20\% of each episode).
Blue: single-agent (TD3); orange: multi-agent (MADDPG).
Dark shades: GRU-equipped; light shades: feedforward.
Lines show rolling mean (window = 10 episodes); shaded envelopes
indicate the range of raw values within each window.}
\label{fig:training_nu}
\end{figure}

Figure~\ref{fig:training_nu} shows how the Nusselt number evolves
over the course of training for all four configurations (single-agent
TD3 and multi-agent MADDPG, each with and without GRU recurrence).
Two metrics are reported. The first is the Nusselt number averaged
over the full episode, $\mathrm{Nu}_{\mathrm{avg}}$; this includes
the initial transient during which the agent is still learning to
modify the flow. The second is the Nusselt number averaged over the
last 20\% portion of the episode, $\mathrm{Nu}_{\mathrm{late}}$; this
reflects how effectively the agent sustains heat transfer reduction
once the flow has fully responded to its actuation.

All four control strategies achieve a similar flow topology
characterised by cell coalescence, reducing $\mathrm{Nu}$ well below
the uncontrolled baseline of 2.48. This confirms that the CNN
architecture is sufficient to learn effective control regardless of
whether a single or multiple agents are used, and stands in contrast
to \citet{Vignon2023}, where achieving coalescence with a
multi-agent approach required re-centring the full field about each
agent, which effectively provides ten times more full-field
trajectories for the shared policy.

The multi-agent strategies consistently reach lower $\mathrm{Nu}$
values than the single-agent approach, particularly in the
late-episode metric, suggesting that distributing control across
local observers better matches the spatially distributed nature of
the convective structures. The GRU-equipped variants learn faster,
most visibly in the multi-agent case where $\mathrm{Nu}_{\text{late}}$
drops below 2.0 within the first 50 episodes, reflecting the benefit
of temporal memory in tracking the delayed response of the flow to
wall temperature changes and anticipating the evolution of the
convective structures. Finally, $\mathrm{Nu}_{\text{avg}}$ proves
more informative than $\mathrm{Nu}_{\text{late}}$ alone, as it
reflects not only the final heat transfer reduction but also how
quickly the flow reorganises into the coalesced state within each
episode: a lower $\mathrm{Nu}_{\text{avg}}$ at comparable
$\mathrm{Nu}_{\text{late}}$ indicates that coalescence is reached
earlier in the episode.

A fixed window of past observations and a larger network are the two obvious
ways of augmenting the baseline feedforward policy without introducing
recurrence, and neither reaches what the recurrent state provides. Both were
trained with every other setting held fixed. Stacking the four most recent
observations as additional input channels gives a median
$\mathrm{Nu}_{\text{late}}$ over three seeds of $1.970$, against $1.865$ with
the GRU and $1.801$ with no temporal input at all, so extending the observation
backwards in time does not improve the converged policy and here degrades it.
Convergence is also markedly slower, the ten-episode rolling mean of
$\mathrm{Nu}_{\text{late}}$ first reaching $1.9$ after 305 episodes against 147
with no temporal input and 49 with the GRU. Widening the actor instead, to
$81{,}537$ trainable parameters against the $79{,}569$ of the recurrent policy,
fares no better: that configuration settles at $2.072$ and does not reach $1.9$
at all within 350 episodes.

Capacity therefore does not account for the difference. The stacked actor
already carries more parameters than the unstacked one, $56{,}337$ against
$54{,}609$, and the widened actor carries more than the recurrent one while
performing worse than either. What distinguishes the recurrent policy is not the
size of its input or of its parameter vector but what it does with them. A stack
of $k$ frames fixes the memory horizon in advance and multiplies the observation
dimension by $k$, whereas the recurrent state learns what to retain and over what
horizon. That matters here because the delay between wall actuation and its
effect on the bulk is set by the diffusive and advective transport times of the
flow rather than by a number available before training.

\begin{figure}[htbp]
\centering
\includegraphics[width=0.95\textwidth]{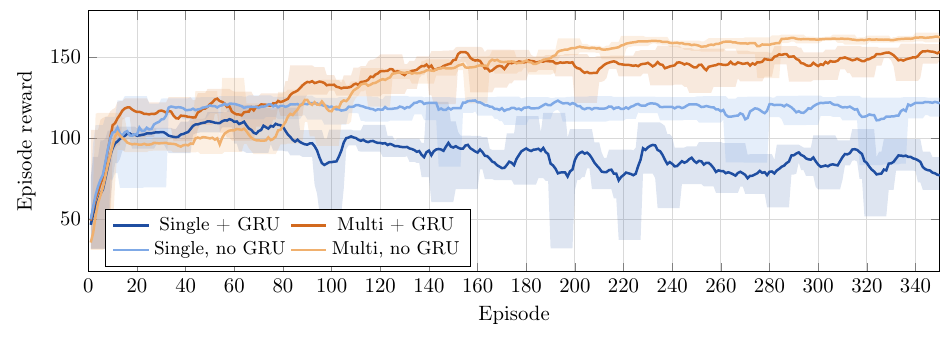}
\caption{Episode reward during training for the four configurations.
Lines show rolling mean (window = 20); shaded envelopes show raw
value range.  Higher reward corresponds to lower $\mathrm{Nu}$.}
\label{fig:training_reward}
\end{figure}

Since the reward is defined as a decreasing function of the
instantaneous Nusselt number, it mirrors the trends described above.
Figure~\ref{fig:training_reward} shows its evolution over training,
confirming that the learning curves are consistent across both
metrics.


\subsection{Comparison with previous results}

Table~\ref{tab:comparison} compares the present approach with
\citet{Vignon2023}. The key difference is architectural:
\citet{Vignon2023} use a flat network that receives the entire flow
field as a vector, discarding its spatial structure, while the
present work uses convolutional networks that process the field as a
two-dimensional array and extract spatially coherent features. In
their multi-agent setup, each agent is given the full flow field
rotated so that its own segment always appears at the centre,
effectively multiplying the number of training trajectories by a
factor of ten. The multi-agent setup proposed here is fundamentally
different: each agent observes only a local patch of 3~neighbouring
segments, with no rotation or data augmentation. The convolutional
architecture achieves comparable results without this augmentation,
indicating that processing the flow field spatially and restricting
each agent to its local neighbourhood are together sufficient to
learn effective control.

\begin{table}[htbp]
\centering
\caption{Comparison with the multi-agent formulation of \citet{Vignon2023}.}
\label{tab:comparison}
\small
\begin{tabular}{p{3.2cm}p{4.0cm}p{4.0cm}}
  \toprule
  \textbf{Aspect} & \textbf{Vignon et~al.} & \textbf{This work} \\
  \midrule
  Algorithm
    & PPO (on-policy)
    & TD3 / MADDPG (off-policy) \\
  Architecture
    & MLP ($2 \times 512$)
    & CNN (stride-2 encoder) \\
  Observation
    & Full field, recentred per agent
    & Full field (single) / local patch (multi) \\
  MARL design
    & Recentring trick (each agent sees full field rotated above it)
    & Local patches (each agent sees 3 neighbouring segments) \\
  Data multiplication
    & 350 ep.\ $\times$ 10 full-field views = 3500 eff.\ trajectories
    & 350 ep.\ (local transitions, no full-field augmentation) \\
  Cell coalescence
    & Yes (MARL only, after 3500 eff.\ trajectories)
    & Yes (all 4 configs in 350 episodes) \\
  Recurrence
    & No
    & GRU (optional, accelerates training) \\
  \bottomrule
\end{tabular}
\end{table}
\subsection{Expected behaviour at higher Rayleigh number}
\label{sec:higherra}

The results above are obtained in a regime where the uncontrolled flow is
steady, and it is worth setting out what would change as $\mathrm{Ra}$ is
raised towards the chaotic regime, since the auxiliary weights of
Table~\ref{tab:aux_losses} are calibrated against physical scales rather
than being free constants.

The temporal weight is referenced to the time the flow takes to respond to a
change in wall temperature, which is governed by conduction across the
thermal boundary layer. That thickness follows $\delta / H \sim 1 /
(2\,\mathrm{Nu})$ and so contracts slowly, from $0.20$ at $\mathrm{Ra} =
10^{4}$ to $0.145$ at $\mathrm{Ra} = 8 \times 10^{4}$, while the ratio of
the diffusive to the free-fall time scale, $\sqrt{\mathrm{Ra}\,\mathrm{Pr}}$,
grows from $84$ to $238$. In free-fall units the response delay is therefore
comparatively insensitive to $\mathrm{Ra}$ and the actuation interval used
here would transfer. What does not transfer is the assumption of a steady
response: once the flow fluctuates on the timescale of the actuation, a
penalty tuned against a steady approach to equilibrium over-regularises.

The spatial weight is referenced to the roll wavelength, and this scale
cannot be inferred from $\mathrm{Ra}$ alone in a domain of fixed width.
Raising $\mathrm{Ra}$ eightfold in the present box in fact \emph{increases}
the dominant horizontal scale, the uncontrolled state passing from four rolls
to two, so the reference length has to be taken from the observed flow. The
same caution applies to the observation resolution, which must continue to
resolve the near-wall gradient as $\delta$ contracts.

Two limitations follow. The controller learnt here exploits a well-defined
topology, so where the flow is chaotic the reward becomes a noisy function of
the actuation and a smaller reduction should be expected. And the reduction
available through coalescence is bounded by how far the uncontrolled state
already sits from the largest structure the domain admits, so results at
different $\mathrm{Ra}$ in a fixed box are not comparable unless that
distance is. We would expect the qualitative conclusion of this work to carry
over to the chaotic regime, but establishing it there is a study in its own
right rather than a change of one parameter.

\section{Evaluation results: Rayleigh--B\'{e}nard convection}
\label{sec:evaluation}

\begin{figure}[htbp]
\centering
\includegraphics[width=0.95\textwidth]{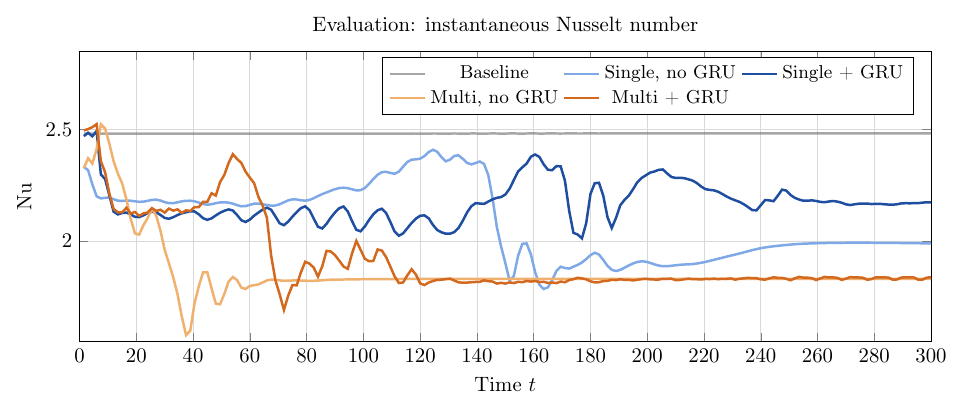}
\caption{Instantaneous Nusselt number during evaluation for all four
configurations. The baseline (grey) maintains $\mathrm{Nu} \approx 2.5$,
while every controlled case reduces $\mathrm{Nu}$ toward the conductive
limit. Blue denotes the single-agent configurations and orange the
multi-agent ones, with dark shades for the GRU-equipped variants.}
\label{fig:eval_nu}
\end{figure}

The four control strategies are evaluated over episodes of 200 actions
(300 free-fall time units $H/U_f$, or approximately 3.6 thermal
diffusion times $H^2/\kappa_T$), starting from the same
uncontrolled steady state used during training. Figure~\ref{fig:eval_nu}
shows the instantaneous Nusselt number during evaluation. The
uncontrolled flow maintains $\mathrm{Nu} \approx 2.5$ throughout,
while all controlled cases progressively reduce $\mathrm{Nu}$ toward
the conductive limit. The multi-agent strategy achieves a deeper and
faster reduction than the single-agent one, consistent with the
advantage observed during training.

\begin{figure}[htbp]
\centering
\includegraphics[width=\textwidth]{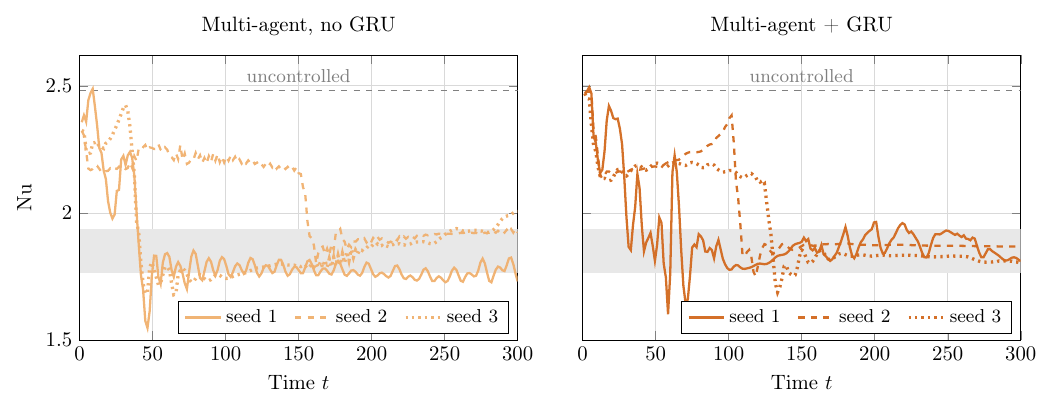}
\caption{Instantaneous Nusselt number during evaluation for the two
multi-agent configurations, each trained three times from independent random
seeds and evaluated from its best checkpoint. Line style identifies the
seed. The grey band spans the plateaux of all six runs, $1.763$ to $1.938$;
the dashed line is the uncontrolled value.}
\label{fig:eval_seeds}
\end{figure}

Individual evaluation curves such as those of Figure~\ref{fig:eval_nu} are
single realisations of a stochastic training process, and the approach to
the controlled state is not a clean transition.
Figure~\ref{fig:eval_seeds} shows the two multi-agent configurations
evaluated from three independent training runs each, differing only in the
random seed. Within one configuration the realisations follow visibly
different paths: one memoryless run reorganises the flow before $t = 50$,
another holds near $\mathrm{Nu} = 2.2$ until $t \approx 150$ and then drops
abruptly, and the third descends gradually. All six settle into the same
interval, $1.763 \le \mathrm{Nu} \le 1.938$, so the level reached by the
multi-agent policies is the same to within the scatter between realisations
and no ordering can be read from any single pair of curves. The recurrent
variant is, however, the more consistent of the two in where it settles: its
three plateaux span $1.817$ to $1.870$, an interval contained entirely
within the one spanned by the memoryless runs, and its curves hold their
level once reached, drifting by $-0.043$, $-0.007$ and $-0.022$ between
mid-episode and the end of the episode where the memoryless runs drift
upward by $+0.001$, $+0.022$ and $+0.139$. The lowest single value of all
six belongs to a memoryless run, so temporal memory is not what sets the
depth of the reduction; what it provides is a policy that reaches a stable
configuration and stays there.

\begin{figure}[htbp]
\centering
\includegraphics[width=\textwidth]{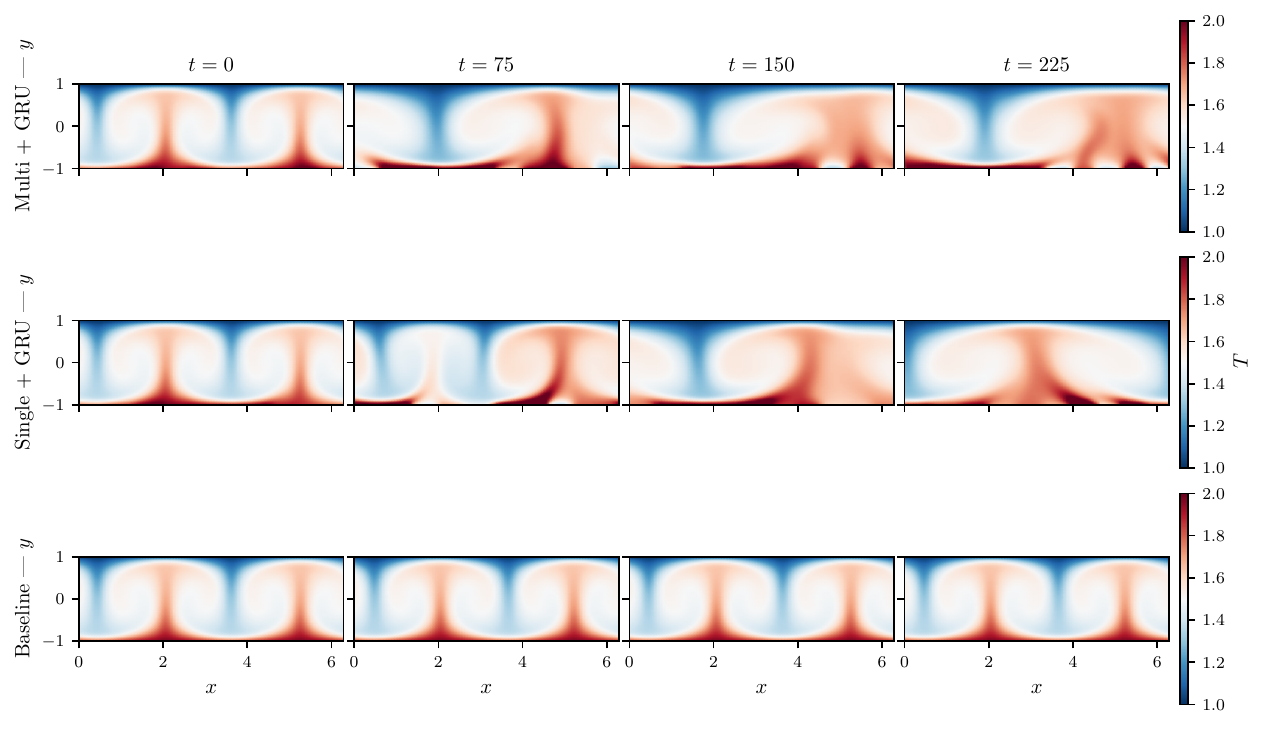}
\caption{Temperature field at four instants during evaluation.
Rows: multi-agent + GRU (top), single-agent + GRU (middle),
uncontrolled baseline (bottom).  The controlled cases show cell
coalescence: merging of convection rolls into larger structures.
The mechanism is apparent at the lower boundary, where the policy
introduces heat at the edges of the base of each cell rather than at
its centre, widening the region held above the mean wall temperature
and so broadening the base over which the cell is supported.}
\label{fig:snap_T}
\end{figure}

\begin{figure}[htbp]
\centering
\includegraphics[width=0.9\textwidth]{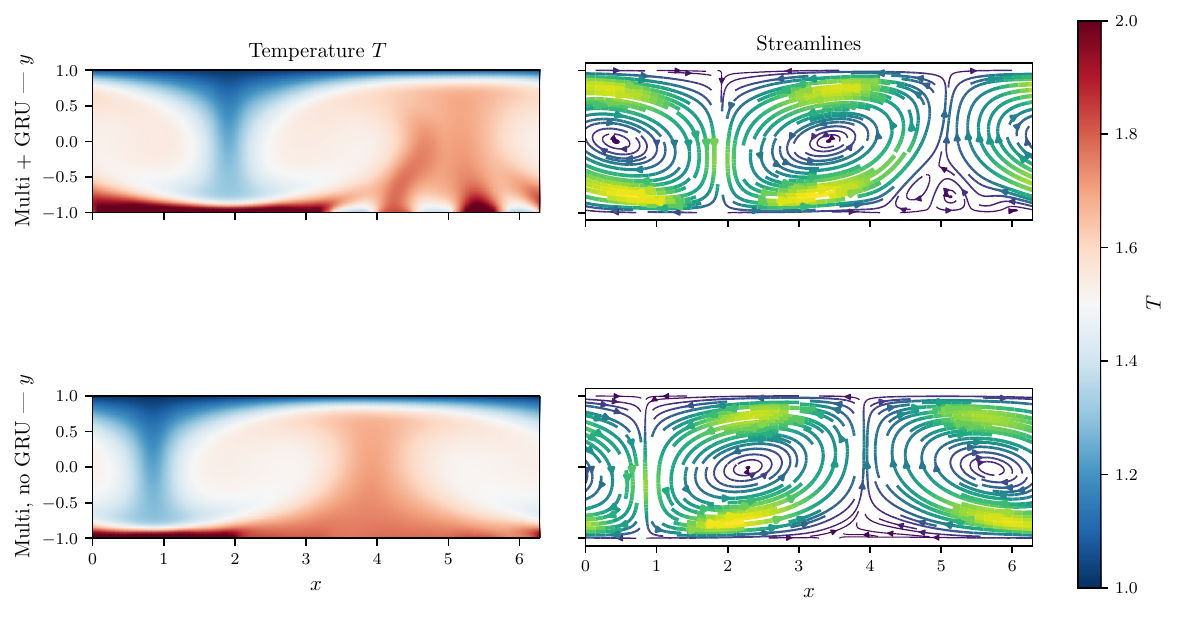}
\caption{Final temperature fields: multi-agent + GRU (top) vs.\
multi-agent without GRU (bottom).  Both reach cell coalescence, but
the GRU variant achieves this earlier during training
(cf.\ Figure~\ref{fig:training_nu}).}
\label{fig:gru_vs_nogru}
\end{figure}

The physical origin of this reduction is visible in
Figure~\ref{fig:snap_T}, which shows the temperature field at
several instants during the episode. All controlled cases undergo
progressive cell coalescence: the initial four-roll structure merges
into fewer, larger rolls, and it is this reorganisation of the
convective topology that drives $\mathrm{Nu}$ downward. The
multi-agent strategy promotes coalescence more rapidly, with the
transition from four to two rolls well under way by $t \approx 75$.
Figure~\ref{fig:gru_vs_nogru} compares the final temperature fields
from the multi-agent strategies with and without GRU memory. Both
reach the coalesced state, confirming that the convolutional
architecture alone provides sufficient spatial expressivity to
discover the correct flow reorganisation. The variant with temporal memory (i.e. the GRU-equipped variant) reaches this state earlier in training
(cf.\ Figure~\ref{fig:training_nu}). 
This is expected, since convective structures take time to reorganise in response to
a change in wall temperature: without temporal context, the same
instantaneous flow snapshot can arise from very different actuation
histories, making it harder to identify an effective control
strategy. The temporal memory resolves this ambiguity by carrying
information about the recent evolution of the flow forward in time.
%

\begin{figure}[htbp]
\centering
\includegraphics[width=0.95\textwidth]{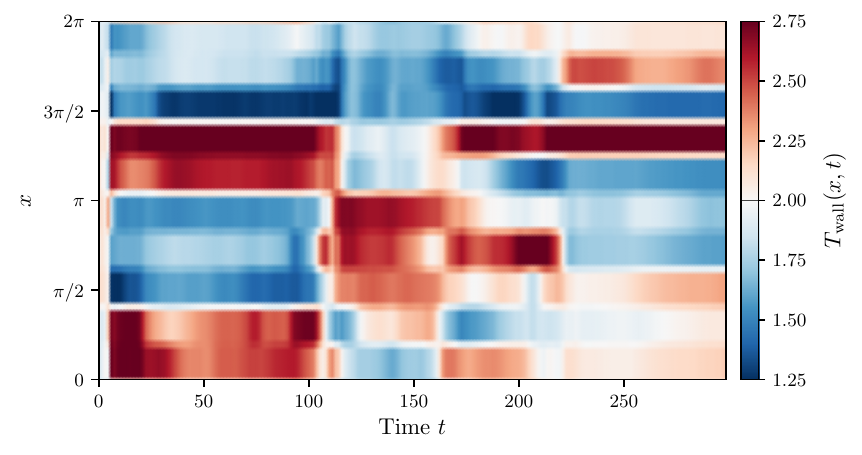}
\caption{Space--time map of bottom-wall temperature $T_{\text{wall}}(x,t)$
for the single-agent + GRU strategy.  The reference value $T_H = 2$
corresponds to white.}
\label{fig:wall_single_gru}
\end{figure}

\begin{figure}[htbp]
\centering
\includegraphics[width=0.95\textwidth]{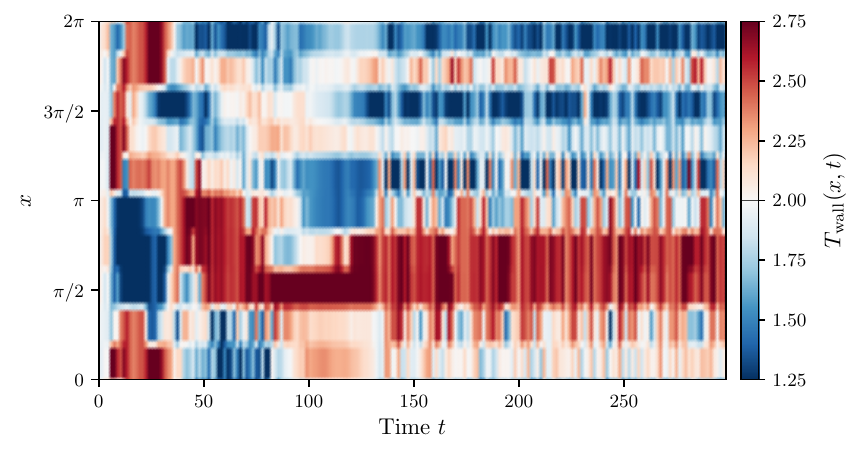}
\caption{Space--time map of bottom-wall temperature for the multi-agent
+ GRU strategy. The edge-heating mechanism of Figure~\ref{fig:snap_T} is
sustained here as a persistently dynamic actuation, the hidden state
carrying information about the preceding evolution of the flow.}
\label{fig:wall_multi_gru}
\end{figure}

\begin{figure}[htbp]
\centering
\includegraphics[width=0.95\textwidth]{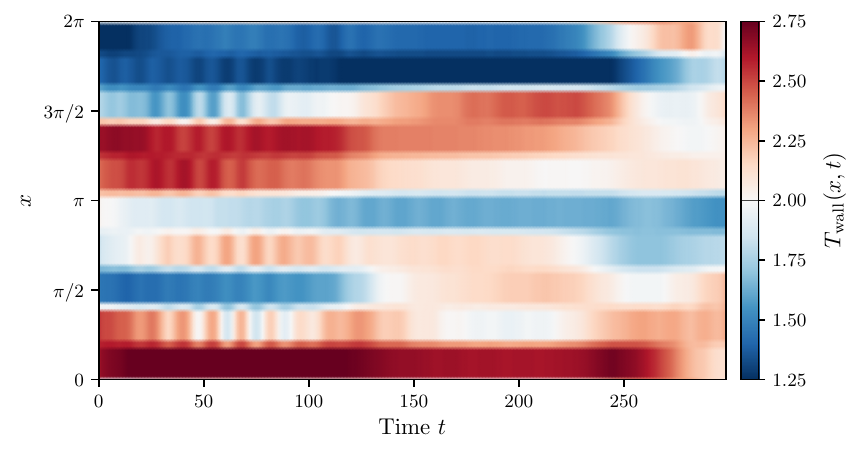}
\caption{Space--time map of bottom-wall temperature for the single-agent,
no GRU strategy.}
\label{fig:wall_single_nogru}
\end{figure}

\begin{figure}[htbp]
\centering
\includegraphics[width=0.95\textwidth]{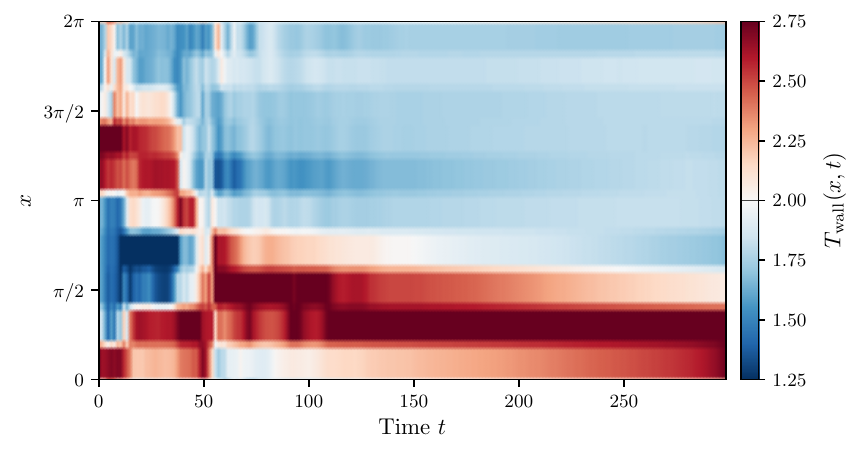}
\caption{Space--time map of bottom-wall temperature for the multi-agent,
no GRU strategy. The same edge-heating mechanism is recovered without
memory, and the actuation is steadier than in
Figure~\ref{fig:wall_multi_gru}, but only once a good configuration has
been reached, at which point the instantaneous field is unambiguous
enough for a memoryless policy to act on.}
\label{fig:wall_multi_nogru}
\end{figure}

The time-averaged Nusselt numbers for all four strategies are
reported in Table~\ref{tab:eval_rbc}. The wall temperature maps in
Figures~\ref{fig:wall_single_gru}--\ref{fig:wall_multi_nogru} show
the spatio-temporal evolution of the actuation applied by each
strategy. In all cases the actuation is smooth and spatially
structured, with no trace of the rapid switching or saturated outputs
that arise when the policy fails to identify a physical control
mechanism. The single-agent maps display broad, slowly evolving
patterns spanning the full domain, while the multi-agent maps exhibit
finer spatial modulation reflecting the individual agents' segment-by-segment control.
The strategies equipped with temporal memory produce smoother
wall-temperature histories than their memoryless counterparts,
reflecting the regularising effect of carrying information forward
across successive actuation steps.

The zero-mean projection of equation~\eqref{eq:zero_mean} guarantees
that the actuation adds no net heat to the bottom wall, but holding a
spatially non-uniform and time-varying wall temperature is not
thermodynamically free, and the engineering value of the control rests
on how its cost compares with the transport it suppresses. A lower
bound on that cost follows from the reversible redistribution of heat
along the wall: transferring a quantity of heat $Q$ from a segment held
at $T_{\text{abs}} - \delta T$ to one held at $T_{\text{abs}} +
\delta T$ requires at least $W_{\text{min}} \simeq 2\,Q\,\delta T /
T_{\text{abs}}$ of work, where $T_{\text{abs}}$ is the absolute
operating temperature. Since the non-dimensional formulation leaves
$T_{\text{abs}}$ free, the ratio must be evaluated at a chosen operating
point. For an air layer with
$\Delta T = 10\,\mathrm{K}$ about an ambient temperature of
$300\,\mathrm{K}$, the largest perturbation admitted by
equation~\eqref{eq:normalise}, $C\,\Delta T = 7.5\,\mathrm{K}$, places
$W_{\text{min}}/Q$ at a few per cent, to be set against the $26\%$
reduction in the heat crossing the layer. This estimate counts
only the reversible work of redistributing heat along the wall, and
excludes the irreversibility of any physical thermal actuator, the
power drawn by sensing and by evaluating the policy, and the transient
cost of establishing the actuation at the start of an episode, all of
which depend on the device rather than on the flow.

\begin{table}[htbp]
\centering
\caption{Evaluation Nusselt numbers (time-averaged over the last 90\%
of the episode) for the four control strategies.}
\label{tab:eval_rbc}
\small
\begin{tabular}{lcc}
  \toprule
  \textbf{Configuration} & $\mathrm{Nu}$ & \textbf{Reduction} \\
  \midrule
  Baseline (uncontrolled) & $\NuBaselineRBCMean$ & --- \\
  Single-agent, no GRU    & $\NuSingleNoGRUMean \pm \NuSingleNoGRUStd$ & $\NuSingleNoGRUReductPct\%$ \\
  Single-agent + GRU      & $\NuSingleGRUMean \pm \NuSingleGRUStd$ & $\NuSingleGRUReductPct\%$ \\
  Multi-agent, no GRU     & $\NuMultiNoGRUMean \pm \NuMultiNoGRUStd$ & $\NuMultiNoGRUReductPct\%$ \\
  Multi-agent + GRU       & $\NuMultiGRUMean \pm \NuMultiGRUStd$ & $\NuMultiGRUReductPct\%$ \\
  \bottomrule
\end{tabular}
\end{table}

\section{Double-diffusive convection in the salt-finger regime}
\label{sec:ddc}

The DRL framework is now extended to double-diffusive convection, where
two scalar fields with different molecular diffusivities, i.e. the temperature $T$ and  the salinity $S$, drive buoyancy simultaneously.
The governing equations retain the Boussinesq framework of Section~\ref{eq:eqs}, augmenting the momentum equation with a two-component buoyancy term and adding a salinity transport equation \citep{turner1974double}:
\begin{align}
  \frac{\partial \bm{u}}{\partial t}
    + (\bm{u} \cdot \nabla)\bm{u}
  &= -\nabla p
    + \sqrt{\frac{\mathrm{Pr}}{\mathrm{Ra}}}\;\nabla^{2}\,\bm{u}
    + \bigl(T - N_{\!\rho}\, S\bigr)\,\hat{\bm{e}}_y,
  \label{eq:dd_momentum} \\[6pt]
  \frac{\partial S}{\partial t}
    + (\bm{u} \cdot \nabla)S
  &= \frac{1}{\mathrm{Le}\,\sqrt{\mathrm{Ra}\,\mathrm{Pr}}}\;\nabla^{2}\, S.
  \label{eq:dd_salinity}
\end{align}
Two additional non-dimensional parameters appear: the Lewis number
$\mathrm{Le} = \kappa_T / \kappa_S$, the ratio of thermal to solutal
diffusivity, and the buoyancy ratio
$N_{\!\rho} = \beta_S\,\Delta S / (\beta_T\,\Delta T)$, which measures
the relative strength of solutal to thermal buoyancy.


These equations admit the salt-finger regime \citep{stern1960salt}, named after the tall,
thin fingers of alternately sinking and rising fluid that form through
a positive feedback: downward-displaced parcels rapidly lose heat but
retain salt, become denser, and continue sinking.
The parameters $\mathrm{Ra} = 7 \times 10^6$, $\mathrm{Pr} = \nu/\kappa_T = 7$
(water-like), $\mathrm{Le} = \kappa_T/\kappa_S = 100$, and
$N_{\!\rho} = 0.167$ are selected to promote this regime; together they
imply a well-separated hierarchy of boundary-layer thicknesses
$\delta_\nu \gg \delta_T \gg \delta_S$, as expected in the salt-finger
regime.
The same $\mathrm{Pr}/\mathrm{Pr}_c$ pair has recently been used in
wall-resolved LES of bounded turbulent double-diffusive convection
across a wider range of thermal and concentration Rayleigh
numbers~\citep{KenjeresRoovers2025}; that work provides a reference
point for the uncontrolled flow physics on which the present control
task acts.
The computational domain is a tall cavity with aspect ratio
$L_x/L_y = 1/3$, resolved on a $128 \times 512$ grid.
Temperature uses Dirichlet boundary conditions ($T_H = 1$ at the
bottom, $T_C = 0$ at the top), while salinity uses no-flux (Neumann)
conditions at both walls.

In this configuration,  the control task differs from the RBC case in both objective and
mechanism: rather than suppressing convection, the goal is to enhance it.
Eight agents ($N = 8$) modulate the bottom-wall temperature in zero-mean
segments, using the same MADDPG framework with GRU-augmented actors
described in Section~\ref{sec:maddpg}. The objective is to
\emph{maximise} the Nusselt number; temperature actuation is a
natural lever in this regime because, with $\mathrm{Le} = 100$, heat
diffuses two orders of magnitude faster than salt, making wall
temperature perturbations an efficient means of modifying the local
buoyancy field.
%

The trained policy is evaluated over a 200-action episode starting
from the uncontrolled steady state.
The controlled policy achieves a sustained increase in heat transfer:
the time-averaged Nusselt number rises from
$\mathrm{Nu}_{\text{base}} = \NuBaseMean \pm \NuBaseStd$ to
$\mathrm{Nu}_{\text{ctrl}} = \NuCtrlMean \pm \NuCtrlStd$, an improvement
of $\NuImprPct\%$.
The Sherwood number $\mathrm{Sh} = \dot{m}_S H / (D_S \Delta S)$, (the solutal analogue of $\mathrm{Nu}$,
measuring the non-dimensional salt flux) decreases slightly by $|\ShImprPct|\%$,
while the salinity spatial variance $\sigma_S^2$ is reduced by
$\SvarReductPct\%$ (from $\SvarBaseMean$ to $\SvarCtrlMean$), indicating
faster mixing. These quantities are summarised in
Table~\ref{tab:dd_results}.
\begin{table}[htbp]
\centering
\caption{Double-diffusive control results (time-averaged over the last
90\% of the episode).}
\label{tab:dd_results}
\begin{tabular}{lccc}
  \toprule
  \textbf{Quantity} & \textbf{Baseline} & \textbf{Controlled}
    & \textbf{Change} \\
  \midrule
  Nusselt $\mathrm{Nu}$
    & $\NuBaseMean \pm \NuBaseStd$
    & $\NuCtrlMean \pm \NuCtrlStd$
    & $+\NuImprPct\%$ \\
  Sherwood $\mathrm{Sh}$
    & $\ShBaseMean \pm \ShBaseStd$
    & $\ShCtrlMean \pm \ShCtrlStd$
    & $\ShImprPct\%$ \\
  Salinity variance $\sigma_S^2$
    & $\SvarBaseMean$
    & $\SvarCtrlMean$
    & $-\SvarReductPct\%$ \\
  \bottomrule
\end{tabular}
\end{table}

\begin{figure}[htbp]
\centering
\includegraphics[width=0.92\textwidth]{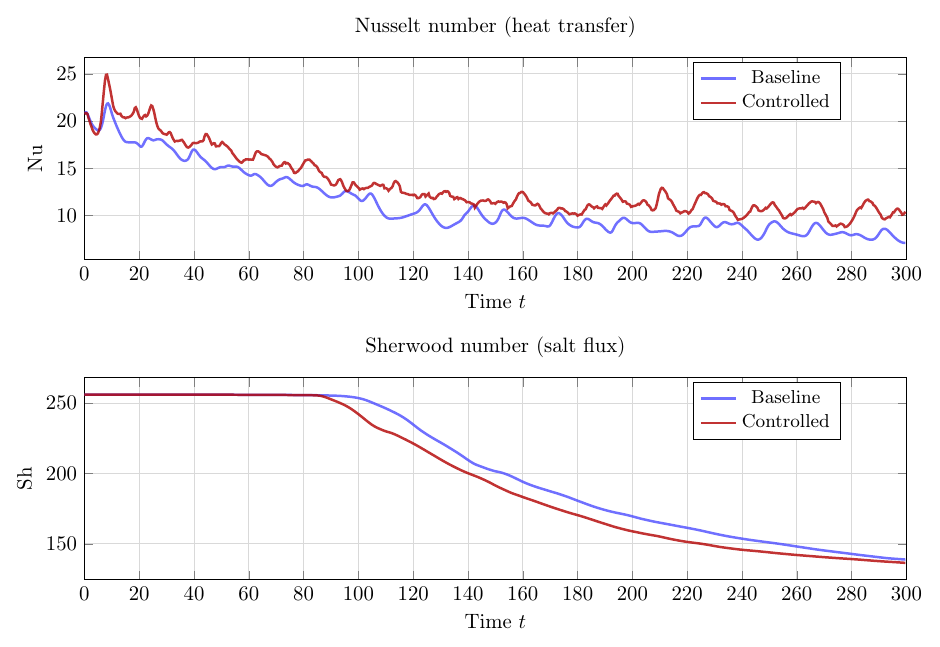}
\caption{Time evolution of $\mathrm{Nu}$ (top) and $\mathrm{Sh}$ (bottom)
for the double-diffusive case. The policy increases $\mathrm{Nu}$ by
$\NuImprPct\%$ while leaving salt flux nearly unchanged.}
\label{fig:dd_nusselt}
\end{figure}

Figure~\ref{fig:dd_nusselt} shows the time evolution of $\mathrm{Nu}$
and $\mathrm{Sh}$. The Nusselt increase is established rapidly and
sustained throughout the episode, while the Sherwood number remains
close to its baseline value, confirming that the policy enhances
thermal mixing without significantly altering the salt flux.

\begin{figure}[htbp]
\centering
\includegraphics[width=0.92\textwidth]{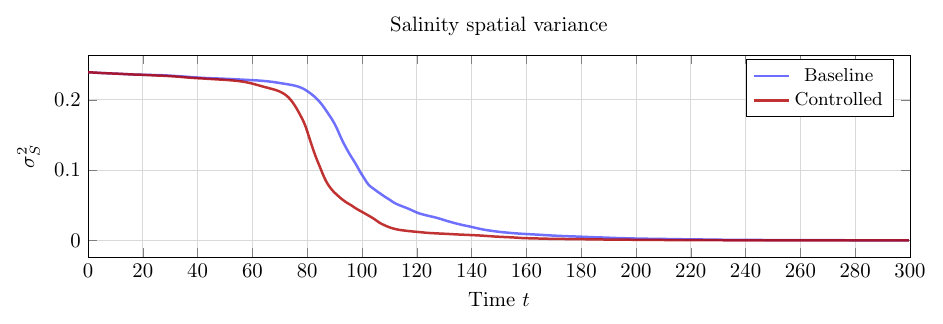}
\caption{Salinity spatial variance $\sigma_S^2$: the controlled case
achieves $\SvarReductPct\%$ reduction relative to the baseline,
reflecting faster mixing of the salt distribution.}
\label{fig:dd_salt_var}
\end{figure}

The evolution of salinity spatial variance $\sigma_S^2$
(Figure~\ref{fig:dd_salt_var}) confirms faster homogenisation: the
controlled case reduces $\sigma_S^2$ by $\SvarReductPct\%$ relative to
the baseline, with the half-life of variance decay shortening from
$\SvarHalfLifeBase$ to $\SvarHalfLifeCtrl$ time units (a
$\SvarSpeedup\times$ speedup).

\begin{figure}[htbp]
\centering
\includegraphics[width=\textwidth]{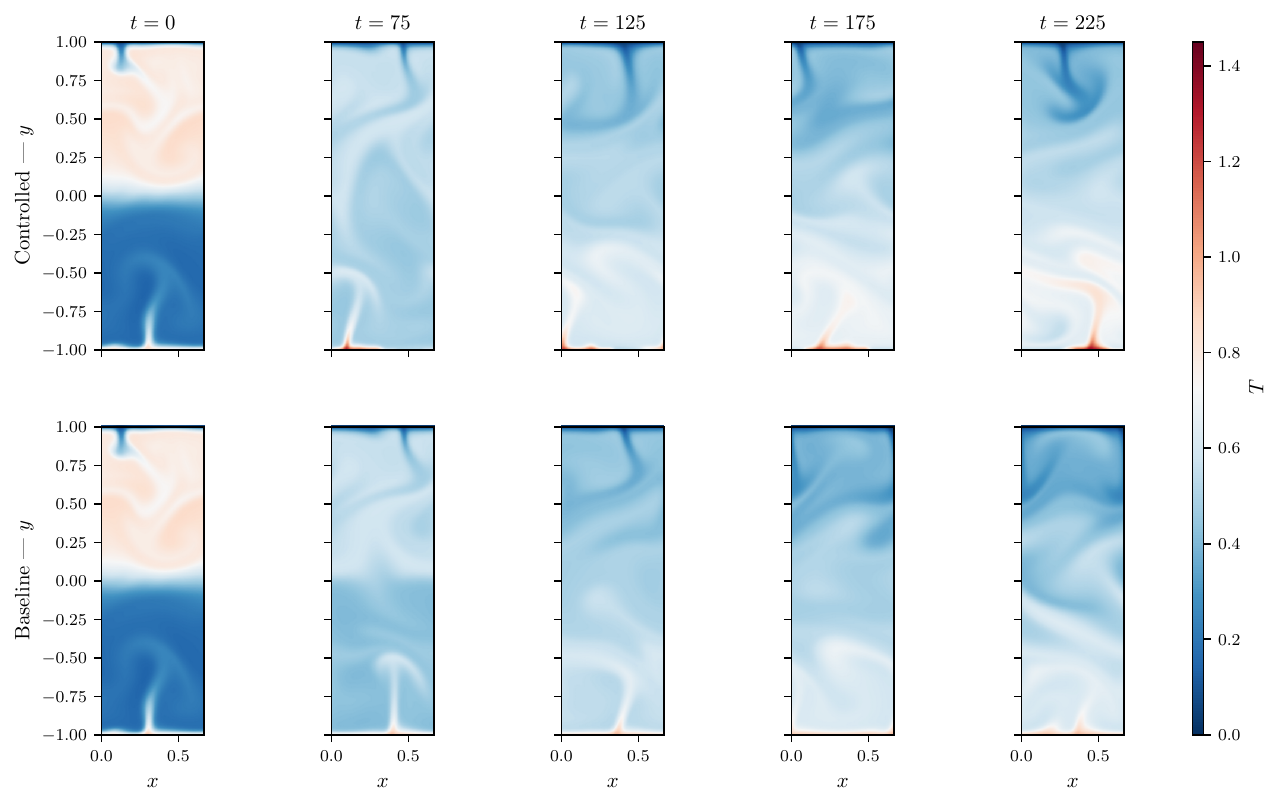}
\caption{Double-diffusive case: temperature field at five instants.
Top row: controlled; bottom row: baseline.}
\label{fig:dd_snap_T}
\end{figure}

\begin{figure}[htbp]
\centering
\includegraphics[width=\textwidth]{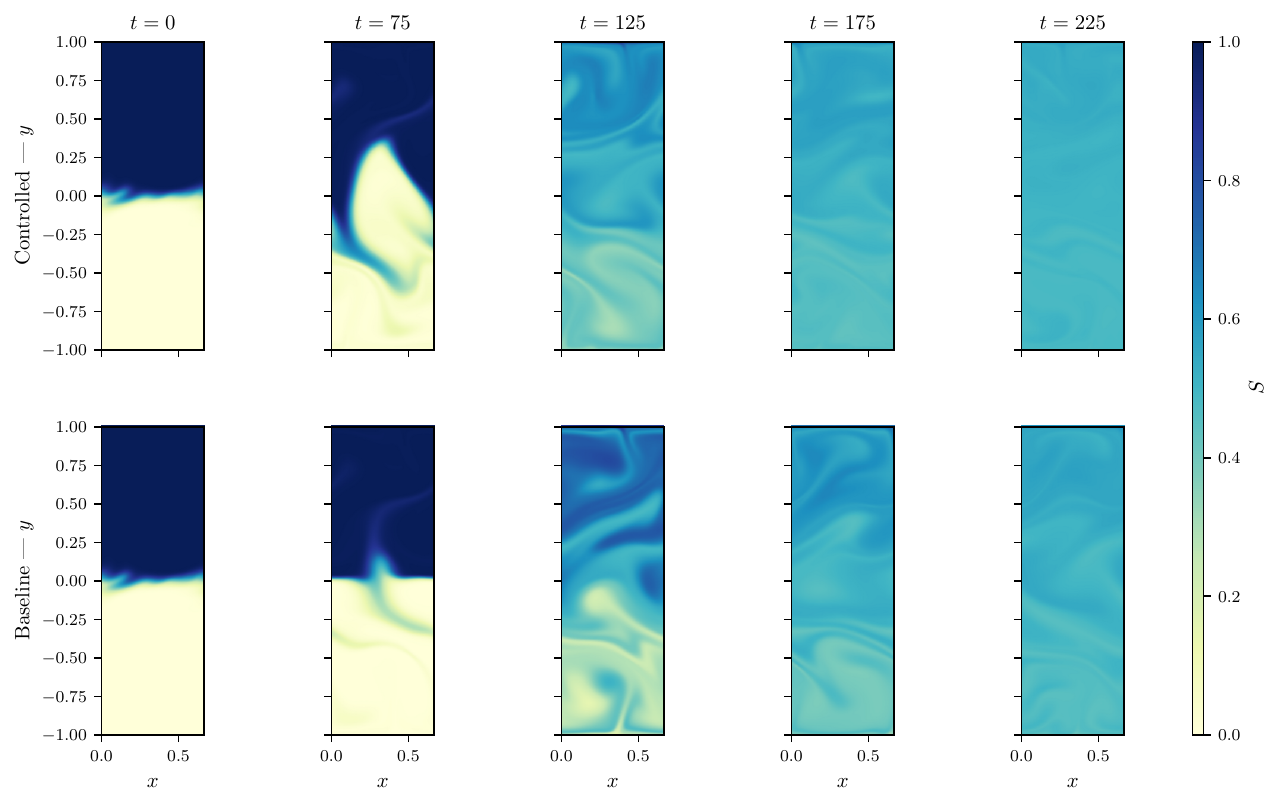}
\caption{Double-diffusive case: salinity field at five instants.
Top row: controlled; bottom row: baseline. The controlled case
shows faster homogenisation of the salt distribution.}
\label{fig:dd_snap_S}
\end{figure}

Figures~\ref{fig:dd_snap_T} and~\ref{fig:dd_snap_S} compare the
temperature and salinity fields at five instants. The controlled case
shows a progressive disruption of the finger pattern and faster
homogenisation of the salt distribution, consistent with the reduced
salinity variance reported above.

%
%

\begin{figure}[htbp]
\centering
\includegraphics[width=0.95\textwidth]{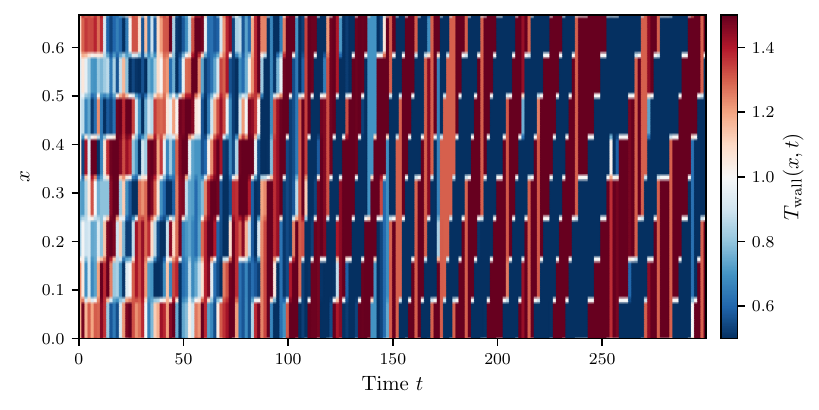}
\caption{Space--time map of bottom-wall temperature for the
double-diffusive case.}
\label{fig:dd_wall}
\end{figure}

\begin{figure}[htbp]
\centering
\includegraphics[width=0.75\textwidth]{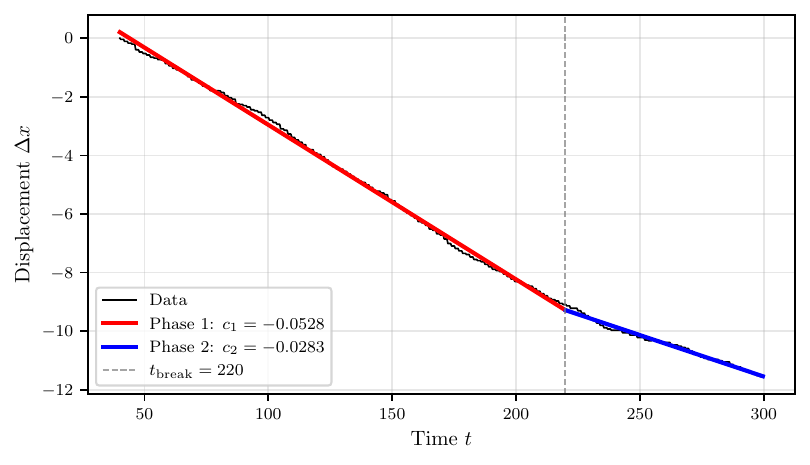}
\caption{Cumulative phase displacement of the $k = 1$ Fourier mode
of the wall temperature perturbation, with piecewise-linear fit.}
\label{fig:dd_phase}
\end{figure}

Beyond the integral quantities, the wall temperature map (Figure~\ref{fig:dd_wall}) reveals a 
more subtle and physically revealing feature of the learned policy:
a coherent travelling-wave structure in the actuation pattern. Tracking the phase
of the fundamental Fourier mode ($k = 1$) of the wall perturbation
yields a piecewise-linear phase displacement
(Figure~\ref{fig:dd_phase}), indicating two distinct regimes of constant
wave speed. During the first phase ($t \lesssim 220$) the pattern
propagates at $c_1 \approx -0.053$, while in the second phase
($t \gtrsim 220$) the wave slows to $c_2 \approx -0.028$, a 46\%
reduction. The slowdown coincides with the salinity field approaching a
more mixed state: as the salt fingers weaken, the policy naturally
reduces the aggressiveness of its stirring strategy. This adaptive
behaviour emerges entirely from the reward signal, without any explicit
encoding of wave speed or mixing state in the control formulation.

\section{Summary and conclusions}
\label{sec:summary}
%
Degenerate actuation, in the form of saturated, pseudo-random, or
spatially incoherent wall-temperature policies, has been identified as
a recurring pathology in DRL-based thermal convection control, rooted
in two compounding deficiencies: MLP policies that discard spatial flow
structure, and memoryless policies that cannot attribute local flow
changes to their own prior actuation. A framework addressing both
causes has been presented, combining convolutional policy networks,
Gated Recurrent Unit (GRU) memory, off-policy training
(TD3\,/\,MADDPG), and explicit action-smoothness constraints. The
framework has been tested on two buoyancy-driven configurations of
increasing physical complexity.

In the Rayleigh--B\'{e}nard convection case at $\mathrm{Ra} = 10{,}000$, all
four agent configurations, single-agent TD3 and multi-agent MADDPG
each with and without GRU recurrence, successfully learn
cell-coalescence strategies, reducing $\mathrm{Nu}$ to as low as
$1.83$ ($26\%$ below the uncontrolled baseline of $2.48$) within
350 training episodes. This matches the effective sample efficiency of
the 3500-trajectory approach of \citet{Vignon2023} without requiring
their re-centred full-field
observations, demonstrating that convolutional inductive biases and
local observations are together sufficient. The multi-agent
configuration yields deeper $\mathrm{Nu}$ reductions than the
single-agent case, reflecting better spectral alignment with the
dominant convective mode, while GRU recurrence accelerates convergence
by approximately 100 episodes across all configurations. Coalescence
is achieved even by the single-agent policy, establishing that the
multi-agent formulation is not a prerequisite once the policy
architecture is sufficiently expressive. The learned actuation in
every case is smooth and spatially structured, with none of the
hallmark failures of degenerate control.

In double-diffusive convection in the salt-finger regime, the
multi-agent recurrent policy enhances heat transfer by $\NuImprPct\%$
and reduces salinity variance by $\SvarReductPct\%$, accelerating
scalar mixing. The learned wall-temperature actuation organises
spontaneously into a coherent travelling wave whose phase speed adjusts
in response to the evolving state of the flow: the wave propagates at
$c_1 \approx -0.053$ during the initial finger-dominated phase and
slows to $c_2 \approx -0.028$ as the salinity field approaches a more
homogeneous state, a $46\%$ reduction in propagation speed. This
state-dependent actuation pattern arises entirely from the scalar
reward signal, with no prescribed wave structure and no explicit
knowledge of the mixing state beyond the instantaneous Nusselt number.
That a boundary-temperature policy should spontaneously select a
travelling-wave strategy to disrupt salt fingers is a physically
non-trivial outcome, one that would be difficult to anticipate or
derive from linear stability arguments alone.

Set against the wider landscape, these results support the view that
the progress reported across the active flow control literature comes
from building the structure of the problem into the formulation rather
than from any particular learning algorithm. \citet{Belus2019} made
that case for the spatial dimension, and multi-agent decomposition has
since been its usual expression; what the present results add is that a
convolutional policy reaches the same end without decomposing the
domain at all, so the decomposition is one way of supplying the
inductive bias rather than the only one. The temporal half of the
problem, identified for wake control by \citet{WangJFM2024} and
addressed there through an explicit delay feature, is met here by
recurrence, which learns the relevant summary of the past instead of
committing to a memory horizon chosen in advance. Read together with
the group-invariant architectures of \citet{Jeon2025}, the picture that
emerges is one of interchangeable mechanisms for injecting the same
spatial and temporal structure, and the practical question for future
work is which of them compound and at what training cost.

Higher Rayleigh numbers will shorten the convective timescale
relative to the actuation period and bring the flow into a chaotic
regime, and it is far from obvious that the recurrent architecture
will remain stable under the much noisier observations that follow.
Three-dimensional geometries add a spanwise dimension to the
convective organisation and a substantial training cost, for which
reduced-order surrogate environments offer a plausible route.
The double-diffusive case hints at a wider class of buoyancy-driven
flows with competing scalar fields, of which thermohaline circulation
and certain crystallisation processes are concrete examples, where
the present framework could be transferred without major modification.
Physical experiments will add sensor noise, actuator inertia and
finite-bandwidth boundary conditions, none of which are represented
in the idealised simulations used here.

\appendix
\setcounter{table}{0}
\section{Sensitivity to the auxiliary loss weights}
\label{app:lambda}

The weights of Table~\ref{tab:aux_losses} are intended to shape the actuation
without reshaping the objective, and it is worth recording how little the
outcome depends on their exact values. Each of $\lambda_{\text{smooth}}$ and
$\lambda_{\text{temporal}}$ was varied by a factor of ten either side of its
tabulated value, all other settings held fixed, on the multi-agent recurrent
configuration. The unmodified configuration was also trained at three random
seeds, which sets the scatter against which any effect has to be judged.
Table~\ref{tab:lambda} reports the average over the final 50 training episodes
of each run.

\begin{table}[htbp]
\centering
\caption{Sensitivity of the trained policy to the smoothness weights. The first
row gives the range obtained from three random seeds at the tabulated weights;
$a_{\text{rms}}$ is the actuation amplitude against a cap of $C = 0.75$.}
\label{tab:lambda}
\small
\setlength{\tabcolsep}{5pt}
\begin{tabular}{llcccc}
  \toprule
  & & $\mathrm{Nu}_{\text{late}}$ & $\langle (a_{j+1}-a_j)^2 \rangle$
  & $\langle (a_j^t - a_j^{t-1})^2 \rangle$ & $a_{\text{rms}}$ \\
  \midrule
  \multicolumn{2}{l}{three seeds, weights as tabulated}
                              &        1.834--2.111 & 0.568--0.687 & 0.012--0.067 & 0.403--0.449 \\
  \midrule
  $\lambda_{\text{smooth}}$   & 0.005  & 1.879 & 0.729 & 0.067 & 0.473 \\
                              & 0.05   & 1.865 & 0.580 & 0.067 & 0.449 \\
                              & 0.5    & 1.904 & 0.245 & 0.022 & 0.404 \\
  \midrule
  $\lambda_{\text{temporal}}$ & 0.0025 & 1.914 & 0.582 & 0.022 & 0.431 \\
                              & 0.025  & 1.865 & 0.580 & 0.067 & 0.449 \\
                              & 0.25   & 1.851 & 0.655 & 0.024 & 0.434 \\
  \bottomrule
\end{tabular}
\end{table}

The control objective is insensitive over the whole range. The late-episode
Nusselt number spans $1.851$ to $1.914$ across two decades of either weight, a
band that lies inside the $1.834$ to $2.111$ obtained from three seeds at the
tabulated values, so either weight can be moved by an order of magnitude with
less effect on the result than changing the random seed produces. The spatial
weight nonetheless acts on the quantity it targets, the roughness falling
monotonically from $0.729$ to $0.245$, although sublinearly, since a
hundredfold change in the weight produces a threefold change in the roughness.
The temporal weight is not resolved by this design, its variation across two
decades moving the realised jitter by less than the seeds move it at fixed
weights. No setting drove the policy towards saturation, the actuation
amplitude remaining between $0.40$ and $0.47$ against the cap of $0.75$
throughout. The penalties therefore guide the optimiser gently, among the many
policies that reduce $\mathrm{Nu}$ comparably well, towards the smooth and
spatially coherent ones, rather than reshaping the objective around themselves.

\setcounter{table}{0}
\section{Sensitivity to the number of wall segments}
\label{app:nseg}

The choice $N = 10$ of Section~\ref{sec:maddpg} was tested by repeating the
multi-agent recurrent training at $N = 5$ and $N = 20$, a factor of four in
segment count, with all other settings held fixed. Table~\ref{tab:nseg} reports
the same quantities as Table~\ref{tab:lambda} together with the number of
episodes taken for the ten-episode rolling mean of $\mathrm{Nu}_{\text{late}}$
to fall to $2.0$, and the range obtained from three random seeds at $N = 10$
again sets the scale against which the differences are read.

\begin{table}[htbp]
\centering
\caption{Sensitivity of the trained policy to the number of wall segments. The
final row gives the range obtained from three random seeds at $N = 10$.}
\label{tab:nseg}
\small
\setlength{\tabcolsep}{5pt}
\begin{tabular}{llccccc}
  \toprule
  $N$ & width & $\mathrm{Nu}_{\text{late}}$ & $\langle (a_{j+1}-a_j)^2 \rangle$
  & $\langle (a_j^t - a_j^{t-1})^2 \rangle$ & $a_{\text{rms}}$
  & episodes to $2.0$ \\
  \midrule
  5   & $1.26H$ & 1.835 & 0.735 & 0.036 & 0.396 & 44 \\
  10  & $0.63H$ & 1.865 & 0.580 & 0.067 & 0.449 & 77 \\
  20  & $0.31H$ & 1.876 & 0.439 & 0.043 & 0.457 & 80 \\
  \midrule
  \multicolumn{2}{l}{three seeds at $N = 10$}
              & 1.834--2.111 & 0.568--0.687 & 0.012--0.067 & 0.403--0.449 & 38--77 \\
  \bottomrule
\end{tabular}
\end{table}

There is no critical segment count within this range. The final Nusselt number
spans $1.835$ to $1.876$ across a fourfold change in $N$, against $1.834$ to
$2.111$ from three seeds at fixed $N$, and the number of episodes needed to
reach $\mathrm{Nu}_{\text{late}} = 2.0$ falls within the $38$ to $80$ spanned by
the same comparison. Coarsening the actuation to five segments, so that a single
segment covers four fifths of an uncontrolled roll, still selects the coalesced
state, and refining it to twenty brings no measurable gain. The decrease of
$\langle (a_{j+1}-a_j)^2 \rangle$ with $N$ follows from measuring the difference
between neighbouring segments on a finer discretisation rather than from a
change in the wall profile, and reflects the actuation approaching a continuous
field as the segmentation is refined, the regime in which the smoothness penalty
rather than the segmentation sets the scale of the actuation.

\section*{Acknowledgements}

G.M.C. acknowledges support from EPSRC UKRI. This work was supported
by the Horizon Europe MSCA DN project with acronym SCALE, Agreement
No.~101120014.

\section*{Code and data availability}

The PyTorch solver, training scripts and post-processing utilities
used to reproduce the results reported here are openly available at
\url{https://github.com/gmcavallazzi/RBC-GRU_MARL}.

\section*{Declaration of competing interest}

The authors declare that they have no known competing financial
interests or personal relationships that could have appeared to
influence the work reported in this paper.

\section*{CRediT author contribution statement}

\textbf{Giorgio Maria Cavallazzi:} Conceptualization, Methodology,
Software, Investigation, Formal analysis, Visualization, Writing --
original draft.
\textbf{Miguel P\'erez Cuadrado:} Software, Data curation, Formal
analysis, Validation, Writing -- review \& editing.
\textbf{Alfredo Pinelli:} Conceptualization, Supervision, Writing --
original draft, Writing -- review \& editing.

\section*{Declaration of generative AI and AI-assisted technologies in the writing process}

During the preparation of this work the authors used Claude
(Anthropic) in order to check grammar and spelling. After using this
tool, the authors reviewed and edited the content as needed and take
full responsibility for the content of the publication.


\end{document}